\documentclass[longauth]{aa}
\usepackage[varg]{txfonts}
\usepackage{float}
\usepackage{longtable}
\usepackage{hyperref}
\hypersetup{
    colorlinks=true,
    citecolor=blue,
    urlcolor=magenta,
    linkcolor=magenta
}
\usepackage{lscape}
\usepackage{amsmath}
\usepackage{subcaption}
\usepackage{color}

\usepackage{subfiles}
\usepackage{booktabs}
\usepackage{placeins}

\usepackage{lineno}
\nolinenumbers

\begin{document}

\newcommand{\rhk}{$\log R'_\text{HK}$}
\newcommand{\cahk}{\ion{Ca}{ii} H\&K}
\newcommand{\ca}{\ion{Ca}{ii}}
\newcommand{\ha}{H$\alpha$}
\newcommand{\nai}{\ion{Na}{i}}
\newcommand{\he}{\ion{He}{i}}

\title{Blind search for activity-sensitive lines in the near-infrared using 
HARPS and NIRPS observations of Proxima and Gl\,581}

\titlerunning{Blind search of activity-sensitive lines in the NIR}

\author{
    Jo\~ao Gomes da Silva\inst{1,*},
    Elisa Delgado-Mena\inst{2,1},
    Nuno C. Santos\inst{1,3},
    Telmo Monteiro\inst{1,3},
    Pierre Larue\inst{4},
    Alejandro Su\'arez Mascare\~no\inst{5,6},
    Xavier Delfosse\inst{4},
    Lucile Mignon\inst{7,4},
    \'Etienne Artigau\inst{8,9},
    Nicola Nari\inst{10,5,6}
% ...
}
        
\institute{
    \inst{1}Instituto de Astrof\'isica e Ci\^encias do Espa\c{c}o, Universidade do Porto, CAUP, Rua das Estrelas, 4150-762 Porto, Portugal\\
    \inst{2}Centro de Astrobiolog\'ia (CAB), CSIC-INTA, Camino Bajo del Castillo s/n, 28692, Villanueva de la Ca\~nada (Madrid), Spain\\
    \inst{3}Departamento de F\'isica e Astronomia, Faculdade de Ci\^encias, Universidade do Porto, Rua do Campo Alegre, 4169-007 Porto, Portugal\\
    \inst{4}Univ. Grenoble Alpes, CNRS, IPAG, F-38000 Grenoble, France\\
    \inst{5}Instituto de Astrof\'isica de Canarias (IAC), Calle V\'ia L\'actea s/n, 38205 La Laguna, Tenerife, Spain\\
    \inst{6}Departamento de Astrof\'isica, Universidad de La Laguna (ULL), 38206 La Laguna, Tenerife, Spain\\
    \inst{7}Observatoire de Gen\`eve, D\'epartement d'Astronomie, Universit\'e de Gen\`eve, Chemin Pegasi 51, 1290 Versoix, Switzerland\\
    \inst{8}Institut Trottier de recherche sur les exoplan\`etes, D\'epartement de Physique, Universit\'e de Montr\'eal, Montr\'eal, Qu\'ebec, Canada\\
    \inst{9}Observatoire du Mont-M\'egantic, Qu\'ebec, Canada\\
    \inst{10}Light Bridges S.L., Observatorio del Teide, Carretera del Observatorio, s/n Guimar, 38500, Tenerife, Canarias, Spain\\
    \inst{11}Instituto de Astrof\'isica e Ci\^encias do Espa\c{c}o, Faculdade de Ci\^encias da Universidade de Lisboa, Campo Grande, 1749-016 Lisboa, Portugal\\
    \inst{12}Departamento de F\'isica da Faculdade de Ci\^encias da Universidade de Lisboa, Edif\'icio C8, 1749-016 Lisboa, Portugal\\
    \inst{13}Departamento de F\'isica Te\'orica e Experimental, Universidade Federal do Rio Grande do Norte, Campus Universit\'ario, Natal, RN, 59072-970, Brazil\\
    \inst{14}Centre of Optics, Photonics and Lasers, Universit\'e Laval, Qu\'ebec, Canada\\
    \inst{15}Herzberg Astronomy and Astrophysics Research Centre, National Research Council of Canada\\
    \inst{16}Aix Marseille Univ, CNRS, CNES, LAM, Marseille, France\\
    \inst{17}Space Research and Planetary Sciences, Physics Institute, University of Bern, Gesellschaftsstrasse 6, 3012 Bern, Switzerland\\
    \inst{18}Center for Space and Habitability, University of Bern, Gesellschaftsstrasse 6, 3012 Bern, Switzerland\\
    \inst{19}Department of Physics, University of Toronto, Toronto, ON M5S 3H4, Canada\\
    \inst{20}Department of Physics \& Astronomy, McMaster University, 1280 Main St W, Hamilton, ON, L8S 4L8, Canada\\
    \inst{21}Department of Physics, McGill University, 3600 rue University, Montr\'eal, QC, H3A 2T8, Canada\\
    \inst{22}Department of Earth \& Planetary Sciences, McGill University, 3450 rue University, Montr\'eal, QC, H3A 0E8, Canada\\
    \inst{23}Centre Vie dans l'Univers, Facult\'e des sciences de l'Universit\'e de Gen\`eve, Quai Ernest-Ansermet 30, 1205 Geneva, Switzerland\\
    \inst{24}European Southern Observatory (ESO), Av. Alonso de Cordova 3107,  Casilla 19001, Santiago de Chile, Chile\\
    \inst{25}Plan\'etarium de Montr\'eal, Espace pour la Vie, 4801 av. Pierre-de Coubertin, Montr\'eal, Qu\'ebec, Canada\\
    \inst{26}Lund Observatory, Division of Astrophysics, Department of Physics, Lund University, Box 118, 221 00 Lund, Sweden\\
    \inst{27}European Southern Observatory (ESO), Karl-Schwarzschild-Str. 2, 85748 Garching bei M\"unchen, Germany\\
    \inst{28}York University, 4700 Keele St, North York, ON M3J 1P3\\
    \inst{29}University of British Columbia, 2329 West Mall, Vancouver, BC, Canada, V6T 1Z4\\
    \inst{30}Western University, Department of Physics \& Astronomy and Institute for Earth and Space Exploration, 1151 Richmond Street, London, ON N6A 3K7, Canada\\
    \inst{31}University Observatory, Faculty of Physics, Ludwig-Maximilians-Universit\"at M\"unchen, Scheinerstr. 1, 81679 Munich, Germany\\
    \inst{32}Hamburger Sternwarte, Gojenbergsweg 112, D-21029 Hamburg, Germany\\
    \inst{33}Subaru Telescope, National Astronomical Observatory of Japan (NAOJ), 650 N Aohoku Place, Hilo, HI 96720, USA\\
    \inst{34}Department of Astronomy \& Astrophysics, University of Chicago, 5640 South Ellis Avenue, Chicago, IL 60637, USA\\
    \inst{35}Consejo Superior de Investigaciones Cient\'ificas (CSIC), E-28006 Madrid, Spain\\
    \inst{36}Bishop's Univeristy, Dept of Physics and Astronomy, Johnson-104E, 2600 College Street, Sherbrooke, QC, Canada, J1M 1Z7\\
    \inst{37}Laboratoire Lagrange, Observatoire de la C\^ote d'Azur, CNRS, Universit\'e C\^ote d'Azur, Nice, France\\
    \inst{38}Department of Physics and Space Science, Royal Military College of Canada, PO Box 17000, Station Forces, Kingston, ON, Canada\\
    \inst{*}\email{joao.silva@astro.up.pt}
}

\date{Received XXX / Accepted XXX}

\abstract{
    Stellar activity variability is one of the main obstacles to the detection of Earth-like planets using the radial velocity (RV) method.
}
{
    The aim of this work is to measure the effect of activity in the spectra of M dwarfs and detect activity-sensitive lines in the near-infrared (NIR) to help improve exoplanet detection and characterisation and contribute to further stellar activity analysis in the NIR.
}
{
    We took advantage of the simultaneous observations of HARPS and the newly commissioned NIRPS spectrograph to carry out a blind search of the most activity-sensitive spectral lines in the NIR using NIRPS spectra and known activity indicators in the optical from HARPS as a reference.
    We analysed the spectra of Proxima (M5.5V) and Gl\,581 (M3V), two M dwarfs with different activity levels and internal structures.
    Spectral lines were identified for both stars and their profiles were fitted using different models.
}
{
    We found hundreds of lines sensitive to activity for both stars; the Proxima spectra were more affected. For Proxima, around 32\% of the identified lines can be used to measure the rotation period of the star, while for Gl\,581 the numbers drops to 1\%.
    The fraction of lines sensitive to activity increases with increasing line depth for both stars.
    A list of 17 lines with rotation period detection for both stars is provided.
}
{
    Stellar activity is able to affect a significant number of spectral lines in the NIR, and methods should be developed to mitigate those effects at the spectral level.
    The line distortions detected here are expected to come mainly from the flux effect due to temperature contrasts between active regions and the quiet photosphere; however, we cannot rule out the possibility that core-emission from chromospheric activity or Zeeman splitting are also affecting some lines.
    The new line lists presented here can be used to improve the RV extraction and the detection of RV variability due to stellar activity signals, and to help false positive detection and the modelling of activity variability, thereby enhancing exoplanet detection in the NIR.
}


\authorrunning{J. Gomes da Silva et al.}

\maketitle

\section{Introduction}

Doppler spectroscopy, also known as the radial velocity (RV) method, has been one of the most successful techniques used in exoplanet detection and characterisation \citep[e.g.][]{mayor2014}.
Due to the advent of extremely high-precision instrumentation such as the Echelle SPectrograph for Rocky Exoplanet and Stable Spectroscopic Observations (ESPRESSO) \citep{pepe2021}, this method is currently capable of detecting planets with RV semi-amplitudes below the 1 m/s level \citep[e.g.][]{faria2022}.
Although instrumental precision has reached the level of $\sim$10 cm/s (similar to the signal caused by Earth orbiting the Sun), this level is particularly difficult to achieve in practice due to the effects caused on RV by stellar variability, or stellar noise, which cannot be reduced with better instrumentation.

Sources of stellar noise can be segregated into different timescales.
Stellar p-mode oscillations are characterised by the propagation of acoustic waves in the convective envelopes of late-type stars, which translates into RV variability \citep[e.g.][]{kjeldsen1995, bedding2001, arentoft2008, kunovac2021}.
In solar-type stars this phenomenon is on timescales of minutes \citep{christensen_dalsgaard2004, kjeldsen2005}, and their effects can be easily attenuated using exposure times with longer timescales or by averaging multiple observations during a given night \citep{dumusque2011a}.
The amplitude of oscillations decreases with decreasing stellar effective temperature, rendering this effect negligible in cooler late-type stars.
Granulation, and its larger counterpart supergranulation, are caused by the convection pattern in the stellar photosphere and have timescales from hours to a few days \citep{de_rosa2004, brandt2008, meunier2015, cegla2018}.
Granulation is also temperature dependent and its influence on RV decreases towards the later-type stars \citep{meunier2017a, meunier2017b, liebing2021}.
The effects of granulation can also be reduced by careful observational strategies \citep{dumusque2011a, meunier2017b}, although the impact of supergranulation on RV is still poorly understood.
Both p-mode oscillations and granulation induce RV variability up to the m/s level \citep[e.g.][]{arentoft2008, bedding2001, meunier2015}.

Stellar activity, in the form of active regions such as spots and faculae, affect RV on timescales typical of the rotation period and its harmonics \citep[][]{saar1997, santos2000, desort2007, lagrange2010, meunier2010, boisse2011, reiners2013}.
The signals depend on the number, the latitudinal and longitudinal distribution of the active regions, and the stellar inclination, making them quasi-periodic \citep[e.g.][]{angus2018} and possibly emulating planetary signals \citep[e.g.][]{bonfils2007, huelano2008, figueira2010, santos2014, robertson2014, carmona2023}.
Activity-induced RV variability can reach hundreds of m/s \citep[e.g.][]{donati2024}, well above the instrumental precision.
Furthermore, the evolution of active regions on timescales of years and decades, in the form of activity cycles \citep{baliunas1995}, can also produce periodic RV variability of the order of tens of m/s in solar-type stars \citep{lovis2011} and M dwarfs \citep{gomesdasilva2012}.

Spectral lines can be affected by stellar activity in different forms, depending on the type of active region, the magnetic field strength, the disk position, and the star spectral type and activity level.
For example, the photometric effect of spots, caused by their temperature contrast to the quiet photosphere, when modulated by stellar rotation, distorts the disk-integrated spectral lines by changing the balance between the blueshifted approaching half of the disk and the redshifted receding half, causing a periodic line asymmetry that affects the centroid position, and therefore the measured RV \citep[e.g.][]{reiners2010,hebrard2014}.
The photometric effect is also known to be wavelength-dependent.
There is evidence that its influence on RV decreases towards redder wavelengths where the spot and faculae contrasts with the photosphere are expected to decrease \citep[e.g.][]{reiners2010, carmona2023}.
However, this dependence on wavelength is primarily driven by contrast, with lower contrasts exhibiting a more pronounced decrease in amplitude with wavelength.

Another effect caused by active regions is the inhibition of convective blueshift.
Convective motions in the photosphere contribute to the granulation pattern where hot plasma rises from the stellar interior in bright granules, cools down, and sinks in the intergranular lanes surrounding the granules.
Since  the areas in the Sun covered by granules are larger than the intergranular lanes, the blueshifted flux from these bright regions is greater than the redshifted flux from the intergranular lanes.
As a result, the overall effect is a net blueshift of the spectral lines \citep{dravins1981}.
This effect affects spectral lines depending on their depth.
Shallower lines that formed in hotter regions deeper in the photosphere are more strongly affected \citep{reiners2016, meunier2017a}.
Active regions are marked by intense magnetic fields \citep[e.g.][]{schrijver2000}, which disable the convection motions and attenuate the convective blueshift.
This disruption alters the balance across the stellar disk, leading to RV variability \citep[e.g.][]{meunier2010}. 
Since the convective blueshift decreases with the effective temperature of stars, this effect is stronger for solar-type stars, but is reduced for M dwarfs \citep{liebing2021}.

Generally, the identification of activity signals is performed using activity-sensitive spectral lines, such as \cahk{}, \ha{}, \nai{}, and others, whose rotational modulation can be analysed using tools such as periodograms \citep[e.g.][]{zechmeister2009_gls, VanderPlas2018}.
The signals observed in these lines can also be used as anchors when modelling the activity contribution to RV variability \citep[e.g.][]{rajpaul2015, faria2020, faria2022}.
Other techniques, such as the cross-correlation function (CCF), can harvest the information of multiple spectral lines into an  average  line.
The parameters of the CCF profile, such as RV, full width at half maximum (FWHM), contrast, and the bisector inverse slope (BIS), are all sensitive to line profile distortions caused by stellar activity \citep[e.g.][]{queloz2001, santos2000, santos2003, santos2010, boisse2011, gomesdasilva2012, figueira2014, dumusque2014}.
However, these methods do not use all the information available in the stellar spectrum.
Although activity-sensitive lines might be more affected by the strong magnetic fields in active regions, all spectral lines are affected by stellar activity variability.
Furthermore, methods that combine spectral lines into an average line profile end up averaging the activity effects on those lines if they affect the lines differently.
Therefore, to better understand the activity effects on RV, it is preferable to include most of the information available in the stellar spectrum by analysing how the activity affects most of the lines \citep{giguere2016}.

Recent studies have used the full spectrum information to analyse how line profiles and their RVs are affected by activity.
Most of them explored the spectra of GK dwarfs, mainly that of alpha Cen B.
\citet{thompson2017} identified line deformations caused by activity;  several lines showed depth variations and FWHM variability modulated by stellar rotation and correlated with the $R'_{HK}$ chromospheric emission ratio.
A list of approximately 40 absorption lines showing line depth correlation with the \cahk{} activity index and rotation modulation was provided by \citet{wise2018} after studying line profile variability.
An extension of this work by \citet{ning2019} identified lines sensitive to activity via an automated method based on Bayesian variable selection using the activity indices based on the \cahk{}, \nai{}, and \ha{} lines and the CCF BIS and FWHM.
Equivalent widths and line asymmetries were measured by \citet{Llsogorskyi2019} and compared to the \cahk{} indicator, who found approximately $350$ lines with strong correlations.
\citet{dumusque2018} showed that by selecting activity-sensitive or -insensitive lines to compute the overall RV, the activity signal could be enhanced or attenuated.
Using line-by-line (LBL) RVs, a relationship between line depth and RV activity sensitivity was detected by \citet{cretignier2020}, where deeper lines show weaker RV activity effects.
This can be explained by the shallower lines closer to the photosphere  that have higher sensitivity to the inhibition of convection caused by active regions.
\citet{al_moulla2022, al_moulla2024} found that measuring RV at different line heights or formation temperatures provides different correlations with activity and different periodicities, meaning that activity influence on RV depends on the line formation temperature.
At higher temperatures, the RV is strongly correlated with inhibition of convection and the \cahk{} index, while at cooler temperatures (higher in the atmosphere) the RV is anti-correlated to the global RV and the \cahk{} index.
Recently, \citet{artigau2024} introduced a method for computing stellar temperature perturbations on individual lines using a library of high-resolution High-Accuracy Radial-velocity Planet Searcher (HARPS) spectra across various spectral types.

In the only study to date focusing on M dwarfs, \citet{lafarga2023} used a similar approach to \citet{dumusque2018}, but applied it to a sample of six early to mid-M dwarfs observed with the visual arm of the high-resolution spectrograph Calar Alto High-Resolution Search for M dwarfs with Exoearths with Near-infrared and Optical Échelle Spectrographs (CARMENES) \citep{quirrenbach2016, quirrenbach2018}.
They found that there is a different activity dependence for the same lines in different stars of similar spectral type, activity level, and rotation period.
They further used LBL RVs insensitive to activity to decrease the RV scatter by around two to five times in their sample stars.

For this work we analysed the influence of activity on the near-infrared (NIR) spectra of two M dwarfs, Gl\,581 and Proxima.
From the perspective of detecting and characterising exoplanets, these cool, small stars offer many advantages over their larger solar-type counterparts.
M dwarfs are abundant, comprising approximately 75\% of all stars in the solar neighbourhood \citep{henry1996}.
This means that most of the planets closer to the solar neighbourhood are orbiting these cool stars.
Their intrinsic low mass makes it easier to detect the RV signals induced by planets, thereby increasing the precision to which lower mass or longer orbit planets can be detected.
Furthermore, their lower luminosity also contributes to a habitable zone closer to the star, enabling the easier detection and characterisation of potentially habitable planets.
Another advantage over FGK dwarfs is that the RV contribution from convective blueshift (and its attenuation by active regions) is very low \citep{beeck2013, Baroch2020, liebing2021} as is the effect of stellar oscillations.
Moreover, activity induced RV amplitude decreases with wavelength, and is expected to be lower at redder and infrared wavelengths \citep{desort2007, reiners2010} where these stars have a higher fraction of their fluxes.

This article is organised as follows. In Sect. \ref{sec:targets} we present the sample stars.
A description of the HARPS and Near-infrared Planet Searcher (NIRPS) data is provided in Sect. \ref{sec:data}.
In Sect. \ref{sec:act} the known activity indicators are analysed along with the detection of the rotation periods.
Our methodology, including the extraction of line profile parameters and the calculation of different pseudo-equivalent widths, is presented in Sect. \ref{sec:methodology} and our results are discussed in Sect. \ref{sec:results}.
We conclude our findings in Sect. \ref{sec:conclusions}.

\section{Target stars}\label{sec:targets}

In this work, our aim is to study the effects of stellar activity and to detect activity-sensitive lines in the spectra of M dwarfs in the NIR regime.
We are interested in finding lines that can be used as activity indicators for stars with different spectral types and activity levels.
This type of analysis is also computer intensive because we  deal  with modelling thousands of spectral lines;  we therefore limited our study to two stars.
We are interested in: (1) stars at the two sides of the limit to fully convective regime, at around spectral type M3--4 \citep{siess2000}, representing two cases of M dwarfs with different internal structures; (2) stars with different activity levels, which we identified visually as having the \ha{} line in emission or absorption; (3) stars with a fair amount of observations and obvious rotation modulations in the activity indicators provided by the Data \& Analysis Center for Exoplanets (DACE),\footnote{\url{https://dace.unige.ch}} namely in the \ha{} index and the FWHM of the NIRPS CCF.

We therefore selected the planet-hosting stars Gl\,581 (M3V) and Proxima (M5.5V), since both present a high number of observations with NIRPS and HARPS, a good signal-to-noise ratio (S/N) in the NIRPS spectra (>100), and a clear rotational modulation signature in the activity indices (see Sect. \ref{sec:act}).
Proxima is a fully convective, active star with emission in both \ha{} and \nai{} lines while Gl\,581 shows absorption in both lines, meaning that its activity level is below that of Proxima.
The H$\alpha$ line is known to first increase in absorption and then in emission as the activity level increases \citep{cram1979,stauffer1986}.
This behaviour could cause confusion when assessing the stellar activity levels.
However, since at high activity levels the emission is above the continuum, we are able to discriminate between high activity levels (strong emission) and low to medium activity levels (absorption).
Nevertheless, to confirm the activity levels of both stars, we estimated their $\log R'_{HK}$ using the \ion{Ca}{ii} H\&K lines extracted from the HARPS spectra and the calibration provided by \citet{suarezmascareno2015}, which includes M dwarfs.
We note that, for M dwarfs, the flux in the \ion{Ca}{ii} H\&K is very low, and after background subtraction, can reach unphysical negative values, which will bias the index towards lower activity levels.
We obtained values of $-5.65 \pm 0.01$ dex and $-5.52 \pm 0.01$ dex for Gl\,581 and Proxima, respectively, showing that Proxima has a higher activity level.
The basic stellar parameters of both stars are presented in Table \ref{tab:star_params}.

\begin{table}
    \tiny
    \centering
    \caption{Stellar parameters for Gl\,581 and Proxima compiled from the literature.}\label{tab:star_params}
    \begin{tabular}{lccccc}
        \hline\hline
        \multicolumn{1}{l}{} & \multicolumn{2}{c}{Gl\,581} & \multicolumn{2}{c}{Proxima} \\
        \hline
        Parameter & Value & Ref. & Value & Ref. \\
        \hline
        V [mag] & 10.560 $\pm$ 0.02 & 1 & 11.13 & 2 \\
        J [mag] & 6.706 $\pm$ 0.026 & 3 & 5.357 $\pm$ 0.023 & 3 \\
        RA [J2000]                            & 15 19 26.83 & 4 & 14 29 42.95 & 4 \\
        DEC [J2000]                           & $-$07 43 20.2 & 4 & $-$62 40 46.2 & 4 \\
        $\mu_\alpha \, \cos{\delta}$ [mas/yr] & $-1221.278$ & 4 & $-3781.741$ & 4 \\
        $\mu_\delta$ [mas/yr]                 & $-97.229$ & 4 & $-20.578$ & 4 \\
        Parallax [mas]                        & 158.718 $\pm$ 0.030 & 4 & 768.067 $\pm$ 0.050 & 4 \\
        Distance [pc]                         & 6.3005 $\pm$ 0.0012 & 4 & 1.3012 $\pm$ 0.0003 & 4 \\
        Spectral Type                         & M3.0\,V & 5 & M5.5\,V & 6 \\
        $L_\star$ [$L_\odot$]                 & 0.01237 $\pm$ 0.00007 & 8 & 0.0016 $\pm$ 0.0006 & 4 \\
        $T_{\mathrm{eff}}$ [K]                & 3367 $\pm$ 91 & 11 & 2900 $\pm$ 100 & {15} \\
        Fe/H                                  & {$-0.21 \pm 0.11$} & {11} & {$-0.18 \pm 0.11$} & {11} \\
        $M_\star$ [$M_\odot$]                 & 0.295 $\pm$ 0.010 & 12 & 0.1221 $\pm$ 0.0022 & 9 \\
        $R_\star$ [$R_\odot$]                 & 0.302 $\pm$ 0.005 & 12 & 0.141 $\pm$ 0.021 & 7 \\
        $P_{rot}$ [days]                      & 132.5 $\pm$ 6.3 & 13 & 83.2 $\pm$ 1.6 & 10 \\
        $P_{cycle}$ [days]                    & $\sim$1400 & 14 & 6560 $\pm$ 85 & 10 \\
        \hline
    \end{tabular}
    \tablefoot{
        References for the provided values: 1 - \citet{zacharias2013}; 2 - \citep{jao2014}; 3 - \citep{cutri2003}; 4 - \citet{GaiaDR12016}; 5 - \citet{reid1995}; 6 - \citet{bessell1991}; 7 - \citet{boyajian2012}; 8 - \citet{cifuentes2023}; 9 - \citet{mann2015}; 10 - \citet{mascareno2025_accepted}; 11 - \citet{antoniadis2024}; 12 - \citet{von_stauffenberg2024}; 13 - \citet{suarezmascareno2015}; 14 - \citet{gomesdasilva2012}; 15 - \citet{pavlenko2017}.
    }
\end{table}

Along with finding spectral lines whose profile variability shows correlations with known activity indices, we are also interested in finding the rotation period of the stars using those lines, and therefore we   need to know their rotation periods.
Several estimates of Gl\,581's rotation period can be found in the literature.
\citet{robertson2014} estimated a rotation period of $130 \pm 2.0$ days for Gl\,581 using the time series of the \ha{} index.
\citet{suarezmascareno2015} used 251 \ion{Ca}{II} H\&K HARPS \citep{mayor2003} observations of this star, with a time baseline of almost 8 years, and obtained a rotation period of \hbox{$132.5 \pm 6.3$ days}.
More recently, \citet{engle2023} found a rotation period of $148.1 \pm 0.9$ days using photometric observations.
In the case of Proxima, \citet{klein2021} detected a rotation period of \hbox{$90 \pm 4$ days} using 10 observations of the longitudinal magnetic field, and the \ha{}, \nai{}, and \he{} activity indices extracted with the HARPS-POL (the extension of HARPS for polarimetry observations) spectropolarimeter \citep{mayor2003, snik2011}.
\citet{faria2022} used Gaussian processes to simultaneously model ESPRESSO \citep{pepe2021} RVs and CCF FWHMs, deriving a rotation period of \hbox{$85.1^{+0.9}_{-0.8}$ days}.
A rotation period of \hbox{$86.4 \pm 2.5$ days} was measured by \citet{engle2023} using photometry.
More recently, \citet{mascareno2025_accepted} measured a rotation period of \hbox{$83.2 \pm 1.6$ days} using the same dataset presented here.

The above rotation periods were derived using different activity proxies, number of observations, time spans, at different epochs, and using different methods, which contribute to the different values obtained by each reference.
We   use these reported rotation periods as reference in our search of rotational variability with the known activity indicators, based on activity-sensitive lines in the visual and the CCF parameters (see Sect. \ref{sec:act}).

\section{Observations and data}\label{sec:data}

\subsection{HARPS and NIRPS spectroscopy}

We used data from the NIRPS spectrograph \cite[][]{bouchy2025_accepted} obtained simultaneously with visual spectra from HARPS \citep[][]{mayor2003}, both installed on the ESO 3.6 m telescope, located at the La Silla Observatory in Chile.
NIRPS is a NIR (0.97--1.81 $\mu$m), fibre-fed, stabilised, high-resolution echelle spectrograph assisted by adaptive optics.
HARPS is an optical (330--690 nm), fiber-fed, stabilised high-resolution echelle spectrograph.

Both Gl\,581 and Proxima were observed as part of Work Package one (WP1) of the NIRPS consortium guarantee time of observation (GTO) under programmes 111.251P.001, 112.25NZ.001 and 112.25NZ.002 (PI: F. Bouchy).
Proxima was observed to demonstrate the potential of NIRPS and to highlight the advantages of using NIR data to search for planets around M dwarfs, and to refine the orbital parameters of its planetary system \citep{mascareno2025_accepted}. Meanwhile, Gl\,581 was selected as a NIRPS target to assess the improvement in RV precision for characterising a planetary system whose central star is known to have an activity signal that is significant for detecting planets in the habitable zone (Larue et al., in prep.).
A total of 82 (Gl\,581) and 137 (Proxima) observations obtained with the high-efficiency (HE) mode (R$\sim$75\,200) were carried out between 2024/04/02 and 2024/09/15 with a median S/N of 156 (Gl\,581) and 187 (Proxima) for NIRPS and 20 (Gl\,581) and 6 (Proxima) for HARPS.
These observations were reduced using an adaptation of the ESPRESSO Data Reduction Software (DRS) pipeline \citep{pepe2021} for NIRPS and HARPS with versions 3.2.0 and 3.0.0 for Gl\,581 and 3.2.0 for both instruments in the case of Proxima.
We used order by order telluric corrected spectra as provided by the DRS.
The CCF was computed by passing the spectra through a weighted cross-correlation \citep{baranne1996, pepe2002} with a numerical mask with a step size equal to instrument pixel width.
The cross-correlation masks used for Gl\,581 were M4V for NIRPS and M3V for HARPS while for Proxima the M5V mask was used for both instruments.
For a description of the telluric correction process, we refer to \citet{allart2022} and \citet{bouchy2025_accepted}.
All spectra were shifted to the stellar rest frame using the CCF RV and the Earth barycentric radial velocity (BERV) provided by the DRS.
The observation log is provided in Table \ref{tab:obs_log}.
All spectra were downloaded from DACE.

\begin{table}
    \small
    \caption{Observation log for Proxima and Gl\,581.}
    \begin{center}
    \begin{tabular}{lccccc}
        \hline\hline
        \multicolumn{1}{l}{} &
        \multicolumn{2}{c}{Gl\,581} &
        \multicolumn{2}{c}{Proxima} \\
        \hline
        \multicolumn{1}{l}{} &
        \multicolumn{1}{c}{HARPS} &
        \multicolumn{1}{c}{NIRPS} &
        \multicolumn{1}{c}{HARPS} &
        \multicolumn{1}{c}{NIRPS} \\
        \hline
        $N_{obs}$ & 80 & 82 & 132 & 137 \\
        $T_{exp}$ [s] & 900 & 451 & 400 & 201 \\
        $\langle S/N \rangle$ & 20 & 156 & 6 & 187 \\
        $\langle S/N_{max} \rangle$ & 63 & 235  & 32  & 290  \\
        $T_{span}$ [d] & \multicolumn{2}{c}{532} & \multicolumn{2}{c}{532} \\
        Start date & \multicolumn{2}{c}{2023--04--02} & \multicolumn{2}{c}{2023--04--02} \\
        End date & \multicolumn{2}{c}{2024--09--15} & \multicolumn{2}{c}{2024--09--15} \\
        \hline
    \end{tabular}
    \end{center}
    \tablefoot{$N_{obs}$ is the number of observations; $T_{exp}$ is the median exposure time; $\langle S/N \rangle$ the median S/N; $\langle S/N_{max} \rangle$ the median maximum S/N; $T_{span}$ is the time span of observations.}
    \label{tab:obs_log}
\end{table}

\subsection{Activity indicators}

Stellar variability caused by active regions can be estimated via the effects they cause on spectral lines by measuring their flux variations or variability in their widths, depths and line bisectors.
To assess the activity variability of our stars, we extracted the CCF parameters FWHM, contrast, and BIS \citep[][]{queloz2001} from the HARPS and NIRPS DRS, along with the activity indices \cahk, \ha06, \ha16, and \nai, using \verb+ACTIN v2.0_beta14+\footnote{\url{https://github.com/gomesdasilva/ACTIN2}} \citep{gomesdasilva2018,gomesdasilva2021}.
The CCF parameters are known to be sensitive to activity variations of cool main sequence FGK \citep{lovis2011}, M \citep{gomesdasilva2012}, and evolved stars \citep{delgadomena2018, delgadomena2023}.
We extracted two versions of the \text{\ha} index, with 0.6 $\mathrm{\AA}$ (\text{\ha06}) and 1.6 $\mathrm{\AA}$ (\text{\ha16}) central bandpasses to compare their behaviour for M-dwarfs.
It was shown that \ha06 is better correlated with the \cahk{} index for FGK dwarfs than \ha16 \citep{gomesdasilva2022}, but there are indications their behaviour is similar for M-dwarfs \citep{meunier2022}.
The \cahk{}, \ha16, and \nai{} indices were analysed in \citet{gomesdasilva2011} for a sample of 30 M-dwarfs using HARPS spectra from the M-dwarf survey \citep{bonfils2007}.
It was shown that \nai{} has a strong correlation with \cahk{}, while \ha16 is only strongly correlated for the more active stars, with the correlation dropping to zero or negative values as the \cahk{} activity level decreases.
As will be discussed in the following section, we ended up not using the \cahk{} indicator in our analysis due to the very low S/N of the HARPS observations, which affects mostly the blue end of the spectrum of M dwarfs, where the \cahk{} are located.

For each time series, a sequential $3\sigma$-clipping from the median for each parameter and respective errors was carried out to remove outliers due to poor telluric correction, weather and instrumental issues, or stellar flares that could complicate our analysis.
All data were binned per day to average out high-frequency variability.

\section{Activity analysis with known indicators}\label{sec:act}

To assess the activity sensitivity of the NIR lines, we need to analyse the known activity indicators behaviour, inter-correlations, and ability to detect the rotation modulation of each star with the available data cadence, time span, and S/N.
By `known indicators' we mean the activity proxies most analysed for M dwarfs, which includes the ones based on the \cahk{}, \ha{}, \nai{}, and \he{} lines \citep[e.g.][]{gomesdasilva2011} and the CCF parameters FWHM, contrast and BIS \citep[e.g.][]{queloz2001, gomesdasilva2012}.
The correlation between these different chromospheric indices and the RV activity signal is studied in more detail in a companion paper by the NIRPS consortium (Larue et al., in prep.).

We used the generalised Lomb-Scargle (GLS) periodogram \citep{zechmeister2009_gls} as implemented by \verb+astropy+ \citep{astropy2022} to search for periods longer than 2.5 days (to exclude daily aliases) and shorter than the time span of observations of each star.
The highest periodogram peaks, more significant than the 0.1\% false alarm probability (FAP) level, and close to the literature rotation periods (see Sect. \ref{sec:targets}) were selected as a rotation period detection.
We do not expect to find exactly the same periods as the reported ones since the data, methods and epochs of detection used are not the same as the ones used here.

Figures \ref{fig:activity_ts_581} and \ref{fig:activity_ts_proxima} show the GLS periodograms for the activity indicators.
The time series will be provided in Larue et al., in prep.
For Proxima, the first season of observations shows a period of 100 days, while in the second season the period is 80.6 days.
This is observed in the GLS, where we have two significant peaks at $\sim$78 and $\sim$98 days surrounding the literature periods. 
\citet{faria2022} showed that the evolution timescale of active regions in Proxima is \hbox{$151^{+38}_{-27}$ days}, a duration corresponding to around two rotation periods.
More recently, as part of the NIRPS consortium and using the same dataset used here, \citet{mascareno2025_accepted} obtained an evolution timescale of \hbox{$58.6^{+1.2}_{-1.2}$ days}, less than the rotation period of the star.
Our data covers 532 days or $\sim$6 rotation periods, so we expect to observe a change in the temporal evolution of active regions during our observing time.
This variability could also contribute to complex behaviour in our activity time series.
Thus, the emergence and disappearance of active regions, at different disk latitudes, can be the cause of the period change observed between the first and second seasons in the time series of this star \citep[see][]{mascareno2025_accepted}.
However, the lower S/N in the HARPS observations for Proxima, with their increased noise, could also explain the double peak observed in the spectral indices for this star.

We found a strong correlation between the NIRPS CCF FWHM and BERV for Gl\,581, with a Pearson coefficient of $\rho = 0.6$.
When comparing the NIR spectral line parameters with the NIRPS CCF FWHM, we need to be careful to identify if the correlation is being driven by BERV (see Sect. \ref{sec:correlations}).
This trend was not observed for Proxima, nor for the other activity indicators.

\subsection{Comparison of activity indices for Gl\,581 and Proxima}

For these M dwarfs, and in agreement with \citet{meunier2022}, the two versions of the \ha{} index have the same behaviour, both detecting similar rotation periods of 77.4 and 77.5 days for Proxima and 140.6 and 139.7 days for Gl\,581, while having Pearson correlation coefficients of $\rho = 0.99$ and $\rho = 0.98$ for each star.
This indicates that the two indices are being affected by the same activity phenomena crossing the stellar disk, contrary to what happens for FGK dwarfs where very often the indices show no correlation or are anti-correlated \citep{gomesdasilva2014, gomesdasilva2022}.
Because of this, and if confirmed for more M dwarfs, the \ha06 should be the \ha{} version of the index used for detecting activity for M and FGK dwarfs.
Using the same bandpass will enable the direct comparison between the activity levels of these two groups of stars (after applying the respective bolometric corrections).
Hereafter we  use the \ha06 index when we refer to \ha{}.

It is known that Gl\,581 exhibits an anti-correlation between the \ha{} and \nai{} emission line \citep[$\rho = -0.75$;][]{gomesdasilva2011} but a positive correlation between \nai{} and the HARPS CCF FWHM \citep[$\rho = 0.60$;][]{gomesdasilva2012}, and we confirm that we observe it in this state.
Concerning Proxima, and consistent with what was observed by \citet{gomesdasilva2011} with the HARPS CCF FWHM, we detect a strong positive correlation between the NIRPS CCF FWHM, \ha{} and \nai{}.
This different behaviour between the indicators could be a consequence of either the activity level \citep[see][]{gomesdasilva2011} or the star internal structure, including the existence of a tachocline for Gl\,581 but not for Proxima, which could be a source of different dynamos operating in the stars resulting in distinctive magnetic field topology in the stellar disks \citep[e.g.][]{siess2000, ossendrijver2003}.

The HARPS CCF parameters of both stars have high noise levels due to the low S/N obtained for this instrument.
This is even more evident in the case of BIS, since its sensitivity decreases for slow rotating stars as shown by \citet{santos2003}.

The fact that we detect slightly different rotation periods for several activity indicators can be due to different activity indices being affected by noise, but also due to different activity phenomena, such as faculae and plages, spots, and filaments \citep[e.g.][]{meunier2009, gomesdasilva2022}, by different amounts.
If these activity phenomena are not spatially correlated in latitude and the star has differential rotation, it is expected that different activity tracers will deliver slightly different rotation periods.

\subsection{Gl\,581 rotation}

According to the discussion in Sect. \ref{sec:targets}, the rotation period is expected to be in the range of 130--148 days.
We detect significant (FAP $\leq$ 0.1\%) periods in this region with several indicators: $P(I_{Ha06}) = 140.6$ days, $P(I_{Ha16}) = 139.7$ days, $P(I_{NaI}) = 139.0$ days, $P(FWHM_{CCF, HARPS}) = 144.8$ days, and $P(FWHM_{CCF, NIRPS}) = 137.4$ days.
Based on the period detection with the activity indices, we chose a broad range of period space to use in our rotation period detection with the NIRPS line parameters.
Our detected periods range between 137.4 and 144.8 days.
To ensure that our search range includes all periods reported in the literature as well as our own detections, with some margin for uncertainty, we decided to limit our range to 130--150 days
This range will be used from now on to identify a rotation period detection for this star.

\subsection{Proxima rotation}

Our literature compilation in Sect. \ref{sec:targets} indicates the rotation period of Proxima is around 80--90 days.
We detect significant periods near this range for several indices: $P(I_{Ha06}) = 77.4$ days, $P(I_{Ha16}) = 77.5$ days, $P(I_{NaI}) = 98.9$ days, $P(FWHM_{CCF, NIRPS}) = 79.0$ days, $P(Cont_{CCF, NIRPS}) = 78.5$ days, and $P(BIS_{CCF, NIRPS}) = 89.7$ days.
There are no detections with HARPS CCF parameters, possibly due to the low S/N for the HARPS data of this star, as shown in Table \ref{tab:obs_log}.
Almost all the indicators that detect a period close to the literature rotation period also show a second peak near 100 days, which in the case of \nai{} is the highest peak.
Although understanding the nature of these two signals is beyond the scope of this study, they can be produced by active regions transiting the stellar disk at different latitudes \citep[see][]{mascareno2025_accepted}.
The periods detected by our activity indicators range between 77.4 and 98.9 days.
Similarly to the case of Gl\,581, we chose a period range capable of including our detected and the reported periods, resulting a range of 70--110 days.
This is the range we use to search for rotational modulation in the line profile parameters.

\section{Methodology}\label{sec:methodology}

Spectral lines were first identified in a high-S/N master spectrum created for each star.
The master spectra was produced as explained in Appendix \ref{sec:master_spectra}.
To identify spectral lines we employed the method of \citet{cretignier2020} as discussed in Appendix \ref{sec:line_id_c20}.
This method also provides a central bandpass for each identified line that we used to integrate $pEW$ and fit the line profiles, as discussed in this section.

To assess line variability, we extracted several parameters of the line profiles.
We extracted parameters using two independent methods to minimise biased results: estimating the line pseudo-equivalent width ($pEW$) and modelling the line profiles to extract the model parameters.
In the next subsections these two methods are explained.

\subsection{Pseudo-equivalent widths}\label{sec:pew}

We calculated the equivalent width of each identified line using two different approaches: (a) a normalisation dependent method, where the pseudo-continuum is assumed to be unity, which we denominate as $pEW_N$ and (b) a normalisation independent method where the pseudo-continuum is estimated at the 0.9 quantile of the flux inside the $2\,\Delta\lambda_{C20}$ (the bandpass extracted via the \citet{cretignier2020} method; see Appendix \ref{sec:line_id_c20}) region centred at the line core, which we designate $pEW_Q$.
Furthermore, and because we do not have an idea of the central bandpass that maximises the activity information that can be extracted from each line, which should be line dependent, we used several central bandpass sizes for each $pEW$ variant.
We calculated five central bandpass sizes by multiplying the bandpass, $\Delta\lambda_{C20}$, by 2, 1, 1/2, 1/3, and 1/4.
Thus, we calculated ten $pEW$ variants, $pEW_{Nk}$ and $pEW_{Qk}$ with $k = 2, 1, 1/2, 1/3, 1/4$ using central bandpasses of $2\,\Delta\lambda_{C20}$, $\,\Delta\lambda_{C20}$, $1/2\,\Delta\lambda_{C20}$, $1/3\,\Delta\lambda_{C20}$, and $1/4\,\Delta\lambda_{C20}$ for the normalisation dependent and normalisation independent cases, respectively.
This methodology will help us avoid not detecting activity variability due to normalisation issues and poor bandpass selection, as well as providing information on the optimal bandpass to measure such variability.

The pseudo-equivalent width of line with centre $\lambda$ for discrete data is given by
\begin{eqnarray}
pEW = \sum^{N}_{i} \left( 1 - \frac{F(\lambda_i)}{F_c} \right) \delta\lambda_i,
\end{eqnarray}
where $N$ is the number of $i$ points (pixels) inside the bandpass $\Delta\lambda$, $F(\lambda_i)$ is the flux of pixel $i$ inside the bandpass, $F_c$ the flux in the pseudo-continuum, and $\delta\lambda_i$ the size of pixel $i$, with $\Delta\lambda = \sum^N_i \delta\lambda_i$.
During the computation of all variants of $pEW$ we performed a cubic interpolation inside the $2\Delta\lambda_{C20}$ region, i.e. the maximum bandpass size considered for each line.
This interpolation was carried out to take into account the finite pixel size, i.e. to align the bandpass edges with the wavelength grid and force the integrated regions inside the bandpass to remain constant for all observations.
A grid step similar to the wavelength step size was used to maintain the wavelength resolution.
The uncertainties in $pEW$ were estimated via error propagation in the flux.

\subsection{Line fitting}\label{sec:line_fitting}

An advantage of fitting the spectral lines over the use of methods such as equivalent widths or core-flux measurements, is that we gain access to line profile parameters such as line centre, width and depth, which provide specific information about which components of the line are being affected by variability.
Deformations of spectral lines by activity phenomena will affect the parameters of the line profile.

We started by fitting all identified lines in the master spectrum\footnote{For the computation of the master spectrum see Sect. \ref{sec:master_spectra}.} using three different functions: Gaussian, Lorentzian, and Voigt.
We used \verb+lmfit+\footnote{\url{https://lmfit.github.io/lmfit-py/}} \citep{lmfit2024} to fit these models to each line, using the flux inside the $\Delta\lambda_{C20}$ central bandpass.\footnote{Obtained from Appendix \ref{sec:line_id_c20}.}
The best model was selected as the one with the lowest $\chi^2_{\mathrm{red}}$ value. The master spectrum fit parameters of the best model, line centre, depth and FWHM were extracted to serve as reference line profile parameters.
These models were then used to fit the observation spectra, and extract time series of the profile parameters for all identified lines.

Before fitting the models, we applied a cubic interpolation to align the pixels with the central bandpass limits.
A wavelength grid with a step similar to the wavelength step of the spectrum was used to maintain the resolution.

\subsection{Correlations with activity}

For each of the line profile parameters and $pEW$ time series obtained from the observations spectra, we calculated the Pearson correlation coefficient, $\rho$, with the three activity indicators $FWHM_{CCF, NIRPS}$, H$\alpha$06, and \ion{Na}{i}.
Significant correlations were chosen as the ones with $|\rho| \geq 0.4$ and $p$--value $\leq 0.01$\%.
This $p$-value was selected to minimise the number of correlations  that can arise `by chance' from calculating hundreds of thousands of correlations.
For example, for Proxima, we identified 2767 lines, have 13 line parameters and three activity indicators, meaning we calculated $2767 \times 13 \times 3 = 107\,913$ correlation coefficients.
For the chosen $p$--value, this means we could have $107\,913 \times 0.0001 \approx 11$ correlations arising `by chance
' (for Gl\,581 this value is lower, since we have a lower number of identified lines).

As mentioned in Sect. \ref{sec:act}, we detected a strong correlation between one of our `anchor' activity indices, $FWHM_{CCF, NIRPS}$, and the NIRPS BERV for Gl\,581.
Correlations between line profile parameters and BERV can arise from telluric contamination or correction residuals affecting the profiles.
In an effort to separate the correlations between the line parameters and activity indicators from those with BERV, we added another criteria for the selection of significant correlations: the correlations between a parameter and activity indices must always be higher that the correlations between the parameter and BERV, $|\rho| > |\rho_{BERV}|$.

\subsection{Rotation period search}\label{sec:prot_search}

Our main interest in this work is to find activity-sensitive lines from an exoplanet detection and characterisation point of view.
Stellar activity generally affects RVs at timescales connected to the rotation period of the star and its harmonics \citep{boisse2011}, and the magnetic cycles periods with timescales that range from years to decades \citep[e.g.][]{baliunas1995, lovis2011, gomesdasilva2012}.
Usually, indicators of activity sensitive to rotation modulation are also able to detect the long-term cycles \citep[e.g.][]{gomesdasilva2022}, since these cycles are characterised mainly by an increase and decrease (along with latitude migration) of the active regions to which the indices are sensitive.
Therefore, by detecting lines sensitive to rotation modulation we are assuming they will also be sensitive to magnetic cycle variations, as is the case of the indices based on the \cahk{}, \ha{}, and \nai{} lines \citep{gomesdasilva2011, gomesdasilva2012, gomesdasilva2014, gomesdasilva2022}.

Another motivation for the period search is that phase shifts between activity indicators can lead to a degradation or a lack of correlations between them \citep{collier_cameron2019} even though the two signals could be following activity variability and be able to detect the stellar rotation.
To account for this, we searched for periodicities in the line parameters that could be close to the star's rotation period (see Sect. \ref{sec:targets}).
We used the GLS periodogram (\citealt{zechmeister2009_gls}; see our Sect. \ref{sec:act}) to search for periodicities in the line parameters time series close to the rotation periods detected with the known indices.
Significant periods were selected as having a GLS power higher than the 0.1\% FAP threshold, computed analytically following \citet{baluev2008}.
We do not report uncertainties on the detected GLS periods because period errors from GLS is a debatable issue \citep[see][]{VanderPlas2018}.
The period search range for each star was the one described in Sect. \ref{sec:act} and the results are presented in Sect. \ref{sec:prot}.

\section{Results}\label{sec:results}

\subsection{Cleaned line list and profile fit accuracy}\label{sec:median_params}

In this section we discuss the line list resulted from the fitting of observations, cleaning of outliers and accuracy of the fitting process, and median distribution of the line profile parameters.

We were able to fit the observations for 2297 lines for Gl\,581 and 3032 lines for Proxima.
Some extracted parameters included very large median errors and dispersions for a few lines that were several orders of magnitude higher than the median values; they  resulted from poor fitting, and we considered them to be outliers.
We tested using $\sigma$ and median absolute deviation (MAD) clipping to clean the data using the median uncertainties and found that applying a 20--MAD clipping from the median was the best solution to remove the outliers for all parameters without the risk of removing too many lines with errors closer to the median of the distribution.
As a result, 63 lines from the Gl\,581 and 44 from the Proxima line lists were removed.
We also performed a similar MAD clip to the median standard deviations, removing a further 127 lines for Gl\,581 and 109 for Proxima.

We compared the median values of the time series from the observations fitting with those obtained from fitting the master spectrum to assess the accuracy of the time series (see Appendix \ref{sec:acc_tests}).
Parameters that deviate significantly from the master fit values indicate disagreement between the profiles in the master and observations spectra (possibly caused by normalisation) or difficulties with the fitting process for the lower-S/N observation spectra.
We are accurate to $\sim$200 m/s in line centre position, $\sim$750 m/s in FWHM, and $\sim$2\% in line depth, at the 1$\sigma$ level.
All lines with a difference between observation and master fit values higher than $\pm$ 3$\sigma$ from the median were removed, affecting 86 lines for Gl\,581 and 112 for proxima.

Finally, our cleaned line list contains 2021 lines for Gl\,581 and 2767 for Proxima, with 1358 lines common to both line lists.\footnote{Common lines were identified when the line centres for one line list were located inside the central bandpass, $\Delta\lambda_{C20}$, for each line in the other line list.}
These are the lines whose parameters we use to compare with the activity indicators and search for rotational modulation.

\subsection{Precision and dispersion of line profile parameters}

\begin{figure*}[h!]
    \centering
        \resizebox{1\hsize}{!}{\includegraphics{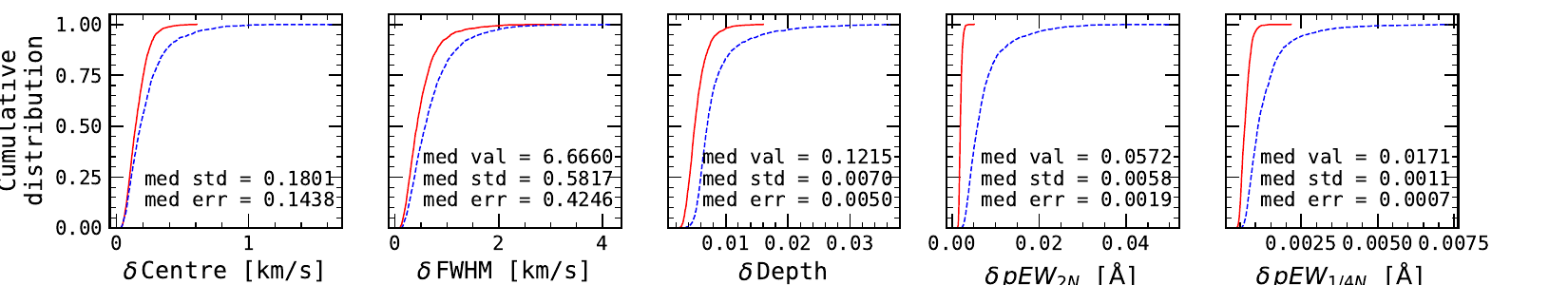}}
    \resizebox{1\hsize}{!}{\includegraphics{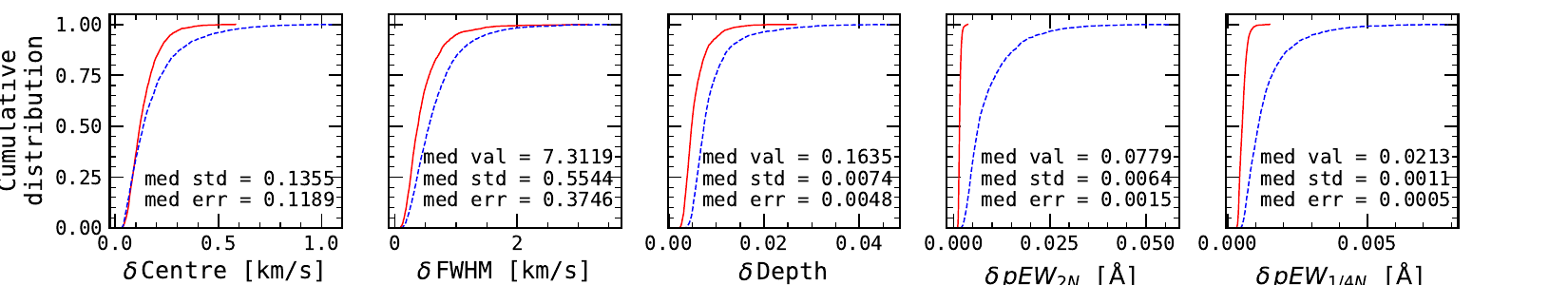}}

        \caption{Cumulative distributions of the median uncertainties and standard deviation for extracted fit parameters and $pEW$ for Gl\,581 (top panels) and Proxima (bottom panels). The red lines are the distribution of uncertainties and the blue lines the distribution of standard deviations. The median value, standard deviation, and error are marked for each parameter.}
        \label{fig:med_errors_std}
\end{figure*}

The distributions of median uncertainties and standard deviations for the extracted fit parameters and $pEW$ are provided in Fig. \ref{fig:med_errors_std}.
Only two of the 10 different $pEW$ variants, one with the wider bandpass ($pEW_{2\mathrm{N}}$) and the other with the narrower bandpass ($pEW_{1/4\mathrm{N}}$), are shown since the distribution of errors and dispersion for the different $pEW$ are very similar.
Overall, we are capable to detect line variability down to the level of the uncertainties of our fitted parameters.
We are able to measure line positions with a median precision of $\sim$130 m/s and median dispersions $\sim$150 m/s.
Our FWHM median uncertainties are at the level of $\sim$400 m/s, with median dispersion of $\sim$600 m/s.
The median precision on the line depths is 0.005, and we measure median dispersions of 0.007.
The minimum dispersion we could measure on individual lines are 27 m/s in line position, 105 m/s in FWHM, and 0.3\% in line depth.
The best precision we could achieve for our lines are 24 m/s in line centre, 71 m/s in FWHM, and 0.2\% in line depth.
This line centre precision is around half of the 50 m/s precision reported by \citet{artigau2022} for the RVs of their best lines, achieved using the LBL technique on Barnard's star with Spectrograph for Planetary Research and Observations (SPiRou), and around two times the precision reported by \citet{dumusque2018} on some spectral lines of $\alpha$ Cen B with HARPS (visual wavelengths) using also the LBL method.
For $pEW$, we obtain median errors between 0.5 and 2 m$\mathrm{\AA{}}$ and median dispersion between 1 and 6 m$\mathrm{\AA{}}$.
This showcases our ability to detect low-amplitude variability in the line profiles of NIR spectra with NIRPS.

\subsection{Correlations with activity indices}\label{sec:correlations}

\begin{figure}
    \centering
        \resizebox{1\hsize}{!}{\includegraphics{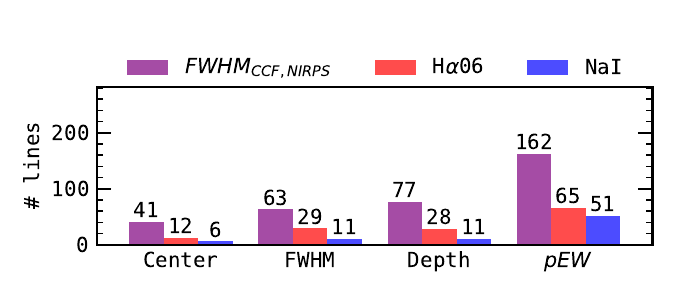}}
    \resizebox{1\hsize}{!}{\includegraphics{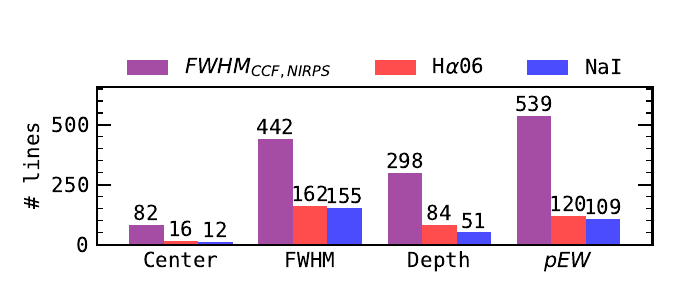}}

        \caption{Number of lines with significant correlations between line profile parameters, $pEW$, and the activity indices for Gl\,581 (top panel) and Proxima (bottom panel).}
        \label{fig:params_correlations}
\end{figure}

\begin{figure*}
    \centering
    \resizebox{\hsize}{!}{\includegraphics{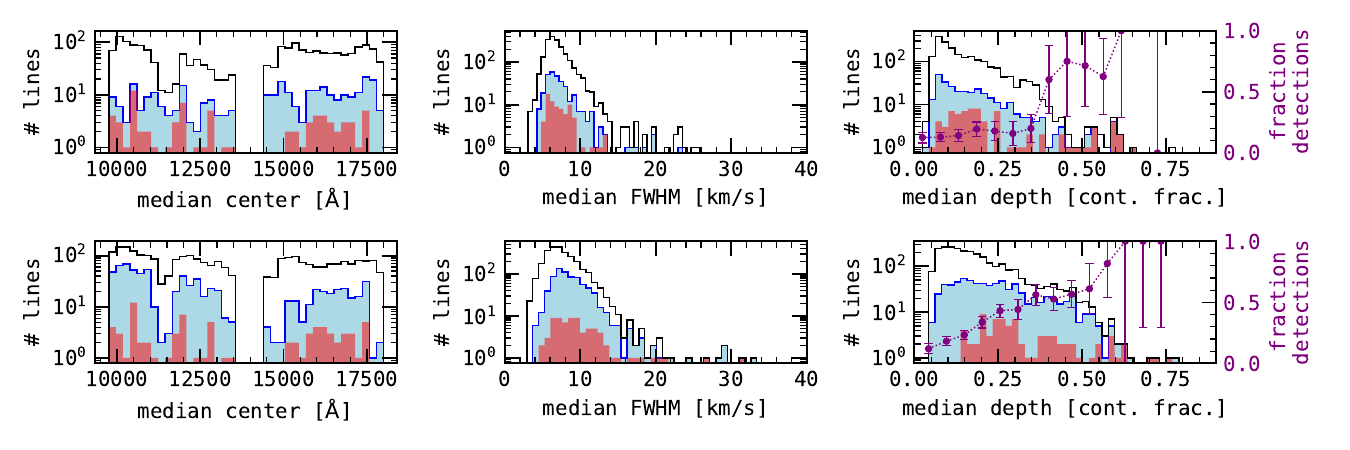}}

    \caption{Distribution of lines with significant correlations ($|\rho| \geq 0.4$, $|\rho| > |\rho_{BERV}|$, and $p$-value $\leq 0.01$\%) with activity in at least one parameter as a function of line profile parameters.
    The black histogram includes all the analysed lines, blue represents the lines with significant correlations between at least one parameter and one activity index, and red shows the lines with significant correlations common to both stars.
    The purple line indicates the fraction of lines with correlations in at least one profile parameter and at least one activity indicator to the full line list; the errorbars are given by Poisson statistics.}
    \label{fig:hist_corr}
\end{figure*}

In this section we compare the number of lines with significant correlations between the line profile parameters, the $pEW$ parameters, and the activity indices.

We found significant correlations between at least one line parameter and one activity indicator for 312 lines (15\%) for Gl\,581 and 760 lines (27\%) for Proxima.
These lines include 78 lines in common between the two stars, which shows that most of the lines with correlations are specific to each star and cannot be used to follow activity, as measured by our indicators, for both stars.

The number of lines with detected significant correlations between our line parameters, line centre, FWHM, depth, and $pEW$, discriminated by the three activity indicators is provided in Fig. \ref{fig:params_correlations}.
The parameter with more detections of significant correlations with activity for both stars is the $pEW$.
This could be a result of the difficulty of fitting and measuring precise small variations in the line profile parameters for lines in the NIR with the NIRPS resolution, S/N, and/or telluric residuals present in these spectra.
But we should also note that the $pEW$ parameter includes 10 different versions that can best be adapted for each line, thus increasing the number of detections.

We can also confirm that all line parameters have higher number correlations with the NIRPS CCF FWHM than the chromospheric activity indices, for both stars.
This can be a result of the different quality of the spectra between NIRPS and HARPS, from where the chromospheric indices were computed, with the S/N of the NIRPS observations being always higher than those of HARPS.
This could also mean that we are mostly detecting line variability that is not caused by chromospheric emission in the lines.
Since we do not expect that the inhibition of convective blueshift has a strong effect on the distortion of lines of M dwarfs \citep{liebing2021}, we could be measuring the flux effect caused by active regions with different temperature contrasts to the quiet photosphere transiting the stellar disk of both stars.
But we could also be measuring the Zeeman splitting of the line profiles caused by the presence of magnetic fields in the active regions \citep[e.g.][]{Fuhrmeister2022}.
For the two stars, the H$\alpha$06 chromospheric indicator always performs slightly better than \ion{Na}{i}, at having correlations with line profile parameters.

In Fig. \ref{fig:hist_corr} we show the distribution of the lines with significant correlation between at least one parameter and one activity index as a function of line centre, FWHM, and depth, for Gl\,581 and Proxima.
Lines with significant correlation in common between the two stars are also shown.
From the distributions of FWHM and line depth, we can observe that Proxima has a larger number of wider and deeper lines than Gl\,581.
In terms of wavelength position, the lines with correlations are in general distributed along all the NIRPS spectral range, although for Gl\,581 there are more of these lines in the $H$ band, redder than 14\,000 $\mathrm{\AA}$, while for Proxima, most of the lines are located in the $Y$ band, between around 10\,000 and 12\,000 $\mathrm{\AA}$.
The lines with significant correlations detected on both stars are also well distributed along the full spectral range.
The distribution of lines with correlations along the FWHM is mostly concentrated in the range around 4--23 km/s, with the median of the distribution at 6.8 km/s for Gl\,581 and 8.3 km/s for Proxima.
The lines with correlations in common between the two stars have a tendency to be slightly wider for Proxima, with the median of the distribution at 9.2 km/s, while for Gl\,581, the median of the distribution stands at 6.8 km/s.
In relation to line depth, in general, the lines of Proxima are deeper than those of Gl\,581, and the lines with significant correlation follow this trend, with the median of the distribution of lines with correlations located at a depth of 0.15 for Gl\,581 and 0.23 for Proxima.
For Gl\,581, the lines with correlations in common for the two stars have a median depth of 0.19, while for Proxima they have a slightly higher median depth of 0.29.
Both stars show increasing fraction of lines with correlations for increasing line depth.
For Gl\,581, the fraction of lines with correlations is almost constant for shallow lines but increases significantly for depths higher than 0.4, reaching around 80\%, while in the case of Proxima there is a steady increase with line depth.
Since most of the lines with detected significant correlations are shallow lines with depths in the range 0.05--0.2, we are confident we are able to measure significant line variability in shallow lines, and that the increasing fraction of correlated lines with line depth is not due to a methodological bias.

\subsection{Rotation period detections}\label{sec:prot}

\begin{figure*}[h!]
    \centering
    \resizebox{\hsize}{!}{\includegraphics{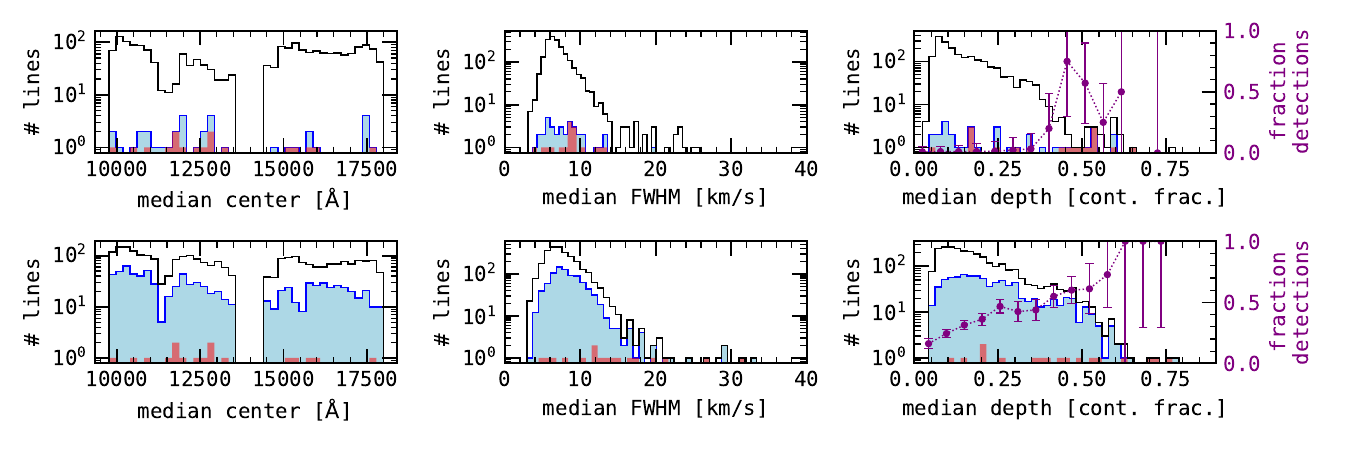}}

    \caption{Distribution of lines with significant $P_{rot}$ detection (FAP $\leq 0.1$\%) as a function of line profile parameters.
    The colours and symbols are the same as in Fig. \ref{fig:hist_corr}.}
    \label{fig:prot_hist}
\end{figure*}

As discussed in Sect. \ref{sec:prot_search}, we carried out a rotation period search using all extracted parameters time series.
Overall, we have 36 lines with significant $P_{rot}$ detection in at least one parameter for Gl\,581 (out of 2021, 1.7\%) and 874 lines (out of 2767, 32\%) for Proxima.
There are 17 of these lines that show profile variability consistent with the rotation period of both stars, which can be used as an activity proxy.
These lines are further discussed in Sect. \ref{sec:final_line_lists}.

The distribution of median line centre, FWHM, and depth for the lines with detected rotation period, for both stars, are provided in Fig. \ref{fig:prot_hist}.
The wavelength distribution of lines with detected $P_{rot}$ is mostly located in the $J$ band for Gl\,581 and $Y$ band for Proxima, with the lines with detections common to the two stars, located mainly in a region between 12\,000-13\,500 $\mathrm{\AA}$.
The lines with rotation detection have FWHM in the region between 4 and 20 km/s for Gl\,581, with the lines of Proxima being more spread in FWHM, between 3 and 33 km/s.
The median values of the FWHM distribution for the lines in common for Gl\,581 and Proxima are 8.8 and 13.6 km/s, respectively.
In terms of line depth, for Gl\,581, the lines with $P_{rot}$ detection appear distributed in two regions, between 0.05 and 0.2 and deeper than 0.4, with the median of the distribution at 0.24.
However, there is a jump in the fraction of lines with detections between 0.4 and 0.6, reaching values of 75\% at around 0.55.
In the case of Proxima, the median of the depth distribution is located at 0.21, and the fraction of lines with a detection increases steadily with line depth, reaching 100\% for depths higher than 0.6, similar to the case of the correlation analysis.

The most striking differences between the lines with significant correlations with activity and the lines with rotation detection is that for Gl\,581 there are only 36 lines with rotation modulation but 312 with correlations with activity, while for Proxima, we found more lines with detected rotation period than lines with significant correlations with activity.

In the case of Gl\,581, our first guess is that, for a line to have detectable rotation modulation on a given parameter, the effects of activity have to have a higher amplitude, than for the line parameters to have correlation with activity, which can vary between low and high correlation coefficients, depending on the effect amplitude on the parameter.
We compared the distribution of fitted parameters standard deviations of line centre, FWHM and line depth for the lines with correlations and the lines with rotation detection.
Contrary to our expectations, the lines with rotation detection are all mainly concentrated at the lowest dispersion range for all parameters ($\sigma$(centre) = 0.04--0.5 km/s; $\sigma$FWHM = 0.18--1 km/s; $\sigma$(depth) = 0.003--0.01), while the lines with correlations with activity range between the full range of the dispersion distribution.
This implies that, having a higher dispersion on the parameters does not ensure that it will be easier to detect the rotation period of the star.

For the case of Proxima, the discrepancy between lines with correlations and lines with rotation detection comes from our thresholds.
Our enforcement that selected correlations must have higher absolute correlation coefficients than the absolute correlation coefficient with BERV is not used for the period analysis, since we can discriminate between the rotation period and periodicities caused by BERV, namely the harmonics of the 1-year signal.
This causes some lines with high correlations with activity not to be selected due to BERV correlation.
Furthermore, we have a very low $p$-value threshold of 0.01\%, which decreases the number of selected lines with correlations.
This is also why we have five lines with detected rotation period for Gl\,581 and 346 for Proxima that do not pass the correlation selection.
The line lists of selected activity-sensitive lines are very sensitive to the imposed selection criteria and different selections can be used, depending on the aims.

Nevertheless, the results described here show that several line profile parameters can be used to measure the rotation period of M dwarfs in the NIR and that, for more active late-type M dwarfs such as Proxima, several hundred lines have their profiles affected by rotational modulation and can be used to infer rotation periods.

\subsection{Activity-sensitive line lists}\label{sec:final_line_lists}

The aim of this work is twofold: to measure the effect of stellar activity on the NIR spectra of M dwarfs and to find new activity-sensitive lines in this wavelength domain.
We identified 2021 lines for Gl\,581 and 2767 for Proxima, which we used to analyse their profile parameters in search for activity effects.

The correlation analysis in Sect. \ref{sec:correlations} provides information about the correlation between different line profile parameters and the three activity indicators $FWHM_{CCF, NIRPS}$, $I_{\mathrm{H}_{\alpha06}}$, and $I_{NaI}$.
The analysis in Sect. \ref{sec:prot} provides information about the periodicities found in the profile parameters.
We deliver these line lists in two CDS tables, one for each star.
A description of columns is provided in Table \ref{tab:table_cds}.
These tables include 312 lines for Gl\,581 and 760 lines for Proxima with significant correlations ($|\rho| \geq 0.4$, $|\rho| > |\rho_{BERV}|$, and $p$--value $\leq 0.01\%$) between at least one line parameter and at least one activity index, and 36 lines for Gl\,581 and 874 lines for Proxima with significant rotation period detection (130 $\leq P_{Gl\,581} \leq 150$ days, \hbox{70 $\leq P_{Proxima} \leq 110$ days}, $\mathrm{FAP} \leq 0.1\%$).
These line lists could be useful to select lines for RV extraction that are not too much affected by activity variability for these two stars.

There are 17 lines with rotation detection, common to both stars, whose basic parameters are provided in Table \ref{tab:prot_paper_table_common}.
To identify the species responsible for each line, we cross-matched the line centres with the VALD3 database \citep{Ryabchikova2015}, selected for the wavelength range of NIRPS and an M-dwarf with the characteristics of Proxima.
In this last table there is a line previously analysed in the literature: \ion{K}{i} $\lambda$12435.644 \citep{Fuhrmeister2022,terrien2022,cortez_zuleta2023}.
Although, in a study involving 324 M dwarfs, \citet{Fuhrmeister2022} found that the \ion{K}{i} line rarely correlates with the strength of the H$\alpha$ line, \citet{terrien2022} found that for Gl\,699 (M4V) and Teegarden's star (M7V), the equivalent width and width of this line is modulated by the rotation of both stars, and that Zeeman splitting could be the cause of such modulation.
In the analysis of Gl\,205 (M1.5V) with SPiRou, \citet{cortez_zuleta2023} did found the rotation period signature in the $pEW$ of \ion{K}{i}, however without enough strength to be considered significant.
Our analysis for Gl\,581 (M3V) and Proxima (M5V) indicate that both the line's FWHM and depth are modulated by rotation.

Our 17 lines with rotational modulation for both stars all have stronger detections for Proxima, as demonstrated by the FAP values in Table \ref{tab:prot_paper_table_common}.
The lines that most significantly detect the rotation of Gl\,581 are \ion{K}{i} $\lambda$12435.644, \ion{Mn}{i} $\lambda$12903.288, \ion{K}{i} $\lambda$11772.858, and \ion{Ti}{i} $\lambda$9790.368, and they should be considered as potential activity indicators in the NIR for M-dwarf stars.

\section{Conclusions}\label{sec:conclusions}

The aim of this study was to analyse the effects of stellar activity in the NIR, and to search for activity-sensitive lines in this wavelength region.
For the first time we identified and analysed for activity effects thousands of NIR lines for two M dwarf stars of the NIRPS GTO with different activity levels and stellar types, Gl\,581 (M3V) and Proxima (M5.5V).

We achieved a median precision on our line profile parameters of $\sim$120 m/s in line centre position, $\sim$400 m/s in FWHM, and $\sim$0.5\% in line depth, at the 1$\sigma$ level. We measured dispersions in the parameters down to the precision level.
We found hundreds of lines with strong correlations with known activity indicators for both stars.
For Gl\,581 15\% of the analysed lines showed strong correlations ($|\rho| \geq 0.4$, $|\rho| > |\rho_{BERV}|$, and FAP $\leq0.01\%$) with the activity indices, while for Proxima we found a higher number of 27\% lines.
These lines can be found along the $Y$, $J$, and $H$ bands, with a tendency for more lines with correlations in the $H$ band for Gl\,581, while for Proxima the tendency is for the majority of the lines to be located in the $Y$ and $J$ bands.
The fraction of lines with correlations tends to increase with line depth for both stars.
Most of the lines with correlations were detected using $pEW$, although we  note that we used ten different versions of this parameter, which can be better adjusted to each line profile, thus contributing  to maximising the number of detections.

We also found hundreds of lines for Proxima capable of detecting the star rotation period (32\%), but only 36 lines for Gl\,581 (1.7\%).
This discrepancy might be a result of the different activity levels of the two stars (Gl\,581 is less active, with no \ha{} emission), but also their internal structure (Proxima with a fully convective interior).

Only a fraction of lines with correlations with activity (78 lines) or rotation period detection (17 lines) are common to both stars, further indicating the spectra are being affected differently for these stars and suggesting that the activity influence on the spectrum of M-dwarf stars should be analysed on a case-by-case basis.

Other results from this work include the following:
\begin{itemize}
    \item Both versions of the \ha{} index have the same behaviour for these M dwarfs, regardless of activity level, spectral type, and internal structure. In the light of the results of \citet{gomesdasilva2022}, we recommend the use of \ha06 for the analysis of FGK and M dwarfs in order to have homogeneous activity indices, which can be used to compare activity between different spectral types (after applying a bolometric correction).
    \item Although the NIRPS CCF FWHM, \ha06, and \nai{} indices are very sensitive to rotation modulation, the first shows a higher number of lines with correlations with line profile parameters and detected rotation periods than the spectral indices based on the \ha{} and \nai{} lines. This is probably due to the higher S/N in the NIRPS observations when compared to HARPS.
\end{itemize}

This work shows that stellar activity affects hundreds of spectral lines in the NIR, mainly at higher activity levels, as is the case of Proxima.
We expect that the activity phenomena distorting the lines arise from the flux effect caused by the different temperature contrasts between active regions and the quiet photosphere, or Zeeman splitting of the line profiles \citep[e.g.][]{Fuhrmeister2022}, although we cannot rule out line-core emission from the chromosphere affecting some lines.
This poses a challenge for exoplanet detection and characterisation using the RV method in the NIR.
Methods for correcting line variability should be developed at the spectral level to mitigate these effects and increase RV precision.

The line lists with the correlation and period detection analysis are provided in CDS for the community.
We curated a list of 17 spectral lines with rotation detection for both stars that can used as potential NIR activity proxies for M dwarfs.
These lists can be used to further develop techniques to analyse and mitigate the effects of stellar activity in the NIR spectra of M dwarfs.
Furthermore, these activity-sensitive lines can be used to reduce the effects of stellar activity on NIR RV by excluding them from CCF masks, LBL analysis, and template matching methods.
Additionally, they can also be used to detect rotation periods and identify false-positive exoplanet detections.

\section*{Data availability}
The tables described in Table \ref{tab:table_cds} are only available in electronic form at the CDS via anonymous ftp to \url{cdsarc.u-strasbg.fr} (130.79.128.5) or via \url{http://cdsweb.u-strasbg.fr/cgi-bin/qcat?J/A+A/}.

\begin{acknowledgements}
    This work has made use of the VALD database, operated at Uppsala University, the Institute of Astronomy RAS in Moscow, and the University of Vienna.
    This publication makes use of The Data \& Analysis Center for Exoplanets (DACE), which is a facility based at the University of Geneva (CH) dedicated to extrasolar planets data visualisation, exchange and analysis. DACE is a platform of the Swiss National Centre of Competence in Research (NCCR) PlanetS, federating the Swiss expertise in Exoplanet research. The DACE platform is available at https://dace.unige.ch.
    JGdS, ED-M, NCS, SCB, ARCS \& EC  acknowledge the support from FCT - Funda\c{c}\~ao para a Ci\^encia e a Tecnologia through national funds by these grants: UIDB/04434/2020, UIDP/04434/2020.
    JGdS and ED-M acknowledge the support from the Funda\c{c}\~ao para a Ci\^encia e a Tecnologia (FCT) through national funds by the grant 2022.04416.PTDC.
    ED-M further acknowledges the support from FCT through Stimulus FCT contract 2021.01294.CEECIND. ED-M  acknowledges the support by the Ram\'on y Cajal contract RyC2022-035854-I funded by MICIU/AEI/10.13039/501100011033 and by ESF+.
    Co-funded by the European Union (ERC, FIERCE, 101052347). Views and opinions expressed are however those of the author(s) only and do not necessarily reflect those of the European Union or the European Research Council. Neither the European Union nor the granting authority can be held responsible for them.
    ASM, NN, JIGH, FGT, VMP, JLR, RR \& AKS  acknowledge financial support from the Spanish Ministry of Science, Innovation and Universities (MICIU) projects PID2020-117493GB-I00 and PID2023-149982NB-I00.
    XDe, XB, ACar, TF \& VY  acknowledge funding from the French ANR under contract number ANR\-18\-CE31\-0019 (SPlaSH), and the French National Research Agency in the framework of the Investissements d'Avenir program (ANR-15-IDEX-02), through the funding of the ``Origin of Life" project of the Grenoble-Alpes University.
    \'EA, RA, FBa, LB, BB, AB, CC, NJC, AD-B, LD, RD, AL, PLam, OL, LMa, LMo, JS-A, PV, TV \& JPW  acknowledge the financial support of the FRQ-NT through the Centre de recherche en astrophysique du Qu\'ebec as well as the support from the Trottier Family Foundation and the Trottier Institute for Research on Exoplanets.
    \'EA, TA, FBa, RD, LMa, J-SM, MO, JS-A \& PV  acknowledges support from Canada Foundation for Innovation (CFI) program, the Universit\'e de Montr\'eal and Universit\'e Laval, the Canada Economic Development (CED) program and the Ministere of Economy, Innovation and Energy (MEIE).
    NN  acknowledges financial support by Light Bridges S.L, Las Palmas de Gran Canaria.
    NN acknowledges funding from Light Bridges for the Doctoral Thesis "Habitable Earth-like planets with ESPRESSO and NIRPS", in cooperation with the Instituto de Astrof\'isica de Canarias, and the use of Indefeasible Computer Rights (ICR) being commissioned at the ASTRO POC project in the Island of Tenerife, Canary Islands (Spain). The ICR-ASTRONOMY used for his research was provided by Light Bridges in cooperation with Hewlett Packard Enterprise (HPE).
    The Board of Observational and Instrumental Astronomy (NAOS) at the Federal University of Rio Grande do Norte's research activities are supported by continuous grants from the Brazilian funding agency CNPq. This study was partially funded by the Coordena\c{c}\~ao de Aperfei\c{c}oamento de Pessoal de N\'ivel Superior—Brasil (CAPES) — Finance Code 001 and the CAPES-Print program.
    KAM acknowledges support from the Swiss National Science Foundation (SNSF) under the Postdoc Mobility grant P500PT\_230225.
    RA acknowledges the Swiss National Science Foundation (SNSF) support under the Post-Doc Mobility grant P500PT\_222212 and the support of the Institut Trottier de Recherche sur les Exoplan\`etes (IREx).
    SCB acknowledges the support from Funda\c{c}\~ao para a Ci\^encia e Tecnologia (FCT) in the form of a work contract through the Scientific Employment Incentive program with reference 2023.06687.CEECIND.
    LB acknowledges the support of the Natural Sciences and Engineering Research Council of Canada (NSERC).
    This project has received funding from the European Research Council (ERC) under the European Union's Horizon 2020 research and innovation programme (project {\sc Spice Dune}, grant agreement No 947634). This material reflects only the authors' views and the Commission is not liable for any use that may be made of the information contained therein.
    This work has been carried out within the framework of the NCCR PlanetS supported by the Swiss National Science Foundation under grants 51NF40\_182901 and 51NF40\_205606.
    BLCM \& AMM  acknowledge CAPES postdoctoral fellowships.
    BLCM acknowledges CNPq research fellowships (Grant No. 305804/2022-7).
    ARCS acknowledges the support from Funda\c{c}ao para a Ci\^encia e a Tecnologia (FCT) through the fellowship 2021.07856.BD.
    NBC acknowledges support from an NSERC Discovery Grant, a Canada Research Chair, and an Arthur B. McDonald Fellowship, and thanks the Trottier Space Institute for its financial support and dynamic intellectual environment.
    LD acknowledges the support of the Natural Sciences and Engineering Research Council of Canada (NSERC) and from the Fonds de recherche du Qu\'ebec (FRQ) - Secteur Nature et technologies.
    JRM acknowledges CNPq research fellowships (Grant No. 308928/2019-9).
    XDu acknowledges the support from the European Research Council (ERC) under the European Union’s Horizon 2020 research and innovation programme (grant agreement SCORE No 851555) and from the Swiss National Science Foundation under the grant SPECTRE (No 200021\_215200).
    DE acknowledge support from the Swiss National Science Foundation for project 200021\_200726. The authors acknowledge the financial support of the SNSF.
    FG acknowledges support from the Fonds de recherche du Qu\'ebec (FRQ) - Secteur Nature et technologies under file \#350366.
    H.J.H. acknowledges funding from eSSENCE (grant number eSSENCE@LU 9:3), the Swedish National Research Council (project number 2023-05307), The Crafoord foundation and the Royal Physiographic Society of Lund, through The Fund of the Walter Gyllenberg Foundation.
    AL  acknowledges support from the Fonds de recherche du Qu\'ebec (FRQ) - Secteur Nature et technologies under file \#349961.
    ICL acknowledges CNPq research fellowships (Grant No. 313103/2022-4).
    LMo acknowledges the support of the Natural Sciences and Engineering Research Council of Canada (NSERC), [funding reference number 589653].
    CMo acknowledges the funding from the Swiss National Science Foundation under grant 200021\_204847 “PlanetsInTime”.
    CPi acknowledges support from the NSERC Vanier scholarship, and the Trottier Family Foundation. CPi  also acknowledges support from the E. Margaret Burbidge Prize Postdoctoral Fellowship from the Brinson Foundation.
    AKS acknowledges financial support from La Caixa Foundation (ID 100010434) under the grant LCF/BQ/DI23/11990071.
    TV acknowledges support from the Fonds de recherche du Qu\'ebec (FRQ) - Secteur Nature et technologies under file \#320056.
\end{acknowledgements}

\bibliographystyle{aa}
\bibliography{bibliography.bib}

\begin{appendix}

\onecolumn
\section{GLS plots of the known activity indicators}

\begin{figure*}[h!]
    \centering
    \begin{subfigure}[t]{0.41\textwidth}

        \includegraphics[width=\linewidth]{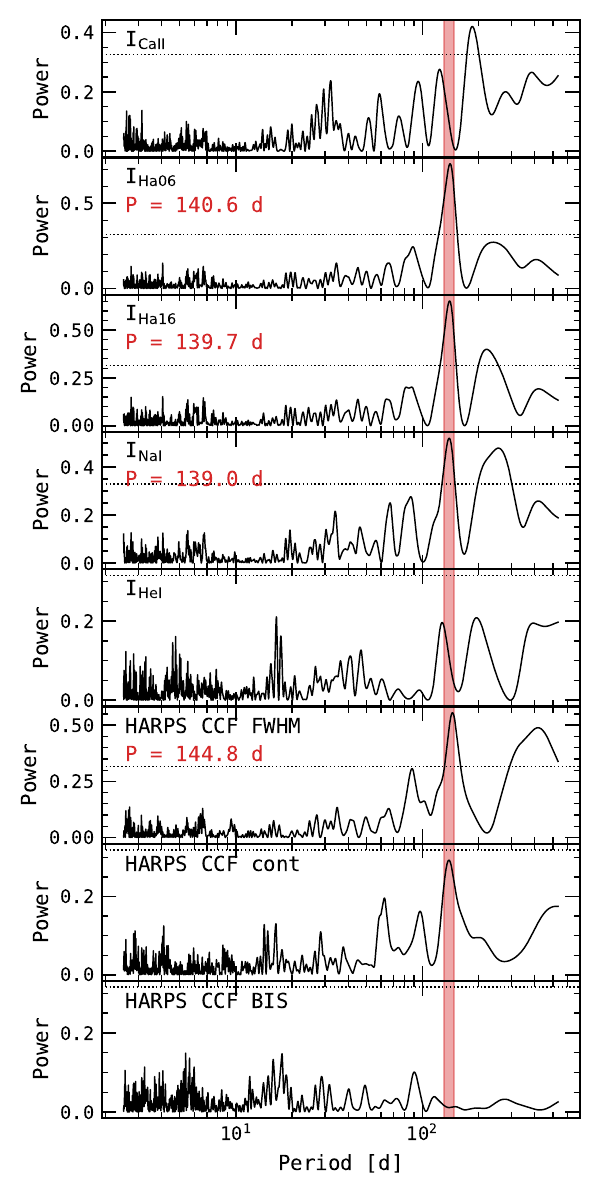}
        \includegraphics[width=\linewidth]{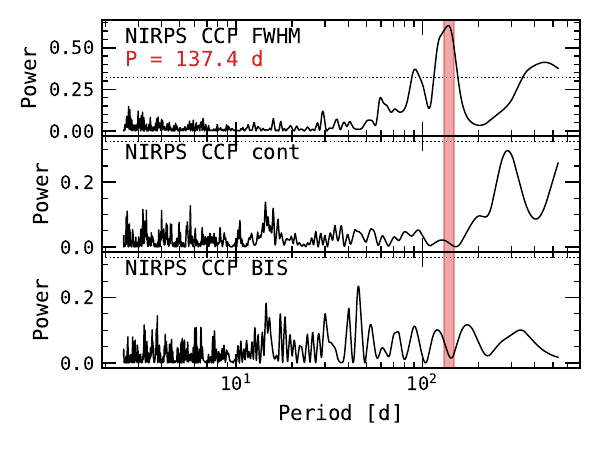}
        \caption{}
        \label{fig:activity_ts_581}
    \end{subfigure}
    \hspace{1cm}
    \begin{subfigure}[t]{0.41\textwidth}
        \includegraphics[width=\linewidth]{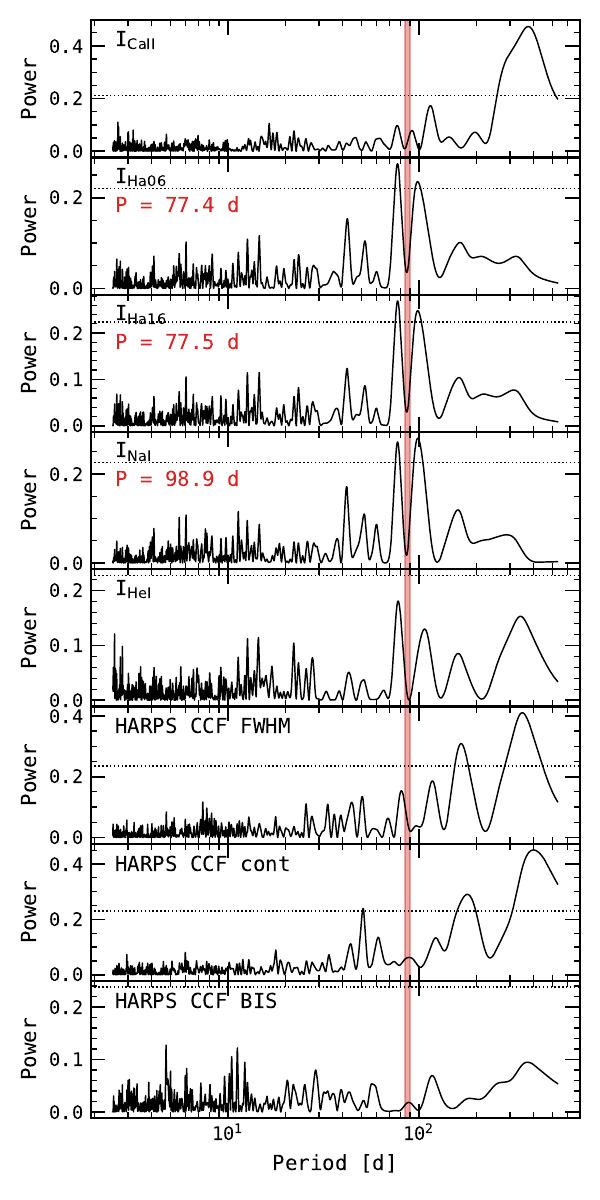}
        \includegraphics[width=\linewidth]{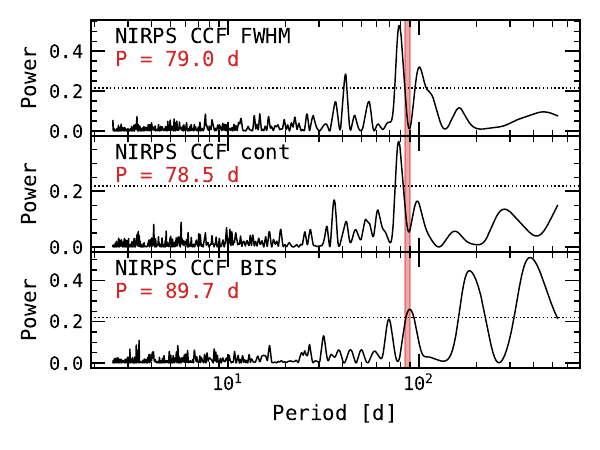}
        \caption{}
        \label{fig:activity_ts_proxima}
    \end{subfigure}
    \caption{HARPS (top panels) and NIRPS (bottom panels) activity index GLS for Gl\,581 (a) and Proxima (b).
    The vertical red band in the GLS marks the literature rotation period span between the minimum and the maximum rotation periods found in the literature.
    The period value in red represents the highest peak closest to the literature values.
    The horizontal dotted lines in GLS indicate the FAP = 0.1\% level}
    \label{fig:activity_ts}
\end{figure*}

\FloatBarrier

\section{NIRPS master spectra}\label{sec:master_spectra}
In this section we describe the computation of the NIRPS master spectra for Gl\,581 and Proxima, which was used to identify the spectral lines to be analysed and to chose the best model, Gaussian, Lorentzian, or Voigt, to fit the individual observations line profiles.

\subsection{Spectra quality selection}

\begin{figure}[t!]
    \centering
    \resizebox{0.5\hsize}{!}{\includegraphics{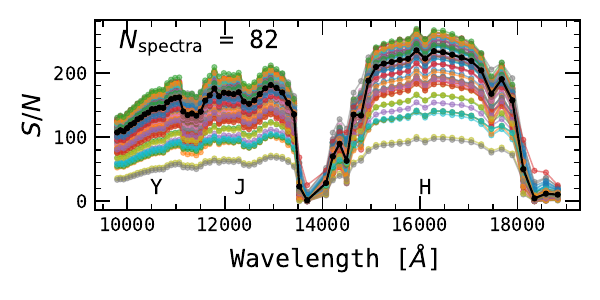}}

        \caption{
                S/N per order for all NIRPS observations of Gl\,581.
        The superposed black line represents the median S/N of all spectra per order.
        The approximate positions of the NIR $Y$, $J$, and $H$ photometric bandpasses are marked above the $x$-axis.
        }
        \label{fig:NIRPS_master_orders}
\end{figure}
\FloatBarrier

Before proceeding to the calculation of the master spectrum, we analysed the quality of the observations to include only high-S/N spectra and spectral orders.
As can be observed in Table \ref{tab:obs_log}, the median S/N of the observations is significantly higher for NIRPS spectra than for HARPS.
For HARPS, all observations with a median S/N below 3 were excluded from the study.
This includes 12 observations for Proxima and one for Gl\,581.
Furthermore, the S/N is not constant for a spectrum along the wavelength range.
Figure \ref{fig:NIRPS_master_orders} shows the S/N per extracted spectral order for all available spectra of Gl\,581.
In the regions between spectral orders 42 and 47 (approximately between 13500 $\AA{}$ and 14100 $\AA()$) and orders higher than 67 ($\gtrsim$18000 $\AA{}$) the S/N quickly drops to values close to zero.
The 42-47 orders region corresponds to the unusable domain of the deep telluric water band between the $J$ and $H$ photometric bands.
After a thorough visual inspection of the individual spectra, we decided to exclude these spectral orders from further analysis in this work.

\FloatBarrier
\subsection{Master spectrum}

To identify the spectral lines to be analysed, we first built a master spectrum, i.e. a high-S/N co-added spectrum comprising all observations, for each star.
When the individual spectra of each observation were shifted to the stellar rest frame by subtracting the stellar systemic and Barycentric Earth RVs, their wavelength grids become slightly different, with different initial and final wavelength values.
Therefore, we need to interpolate all spectra into a common wavelength grid to be able to co-add them.
To construct the new grid, we determined each spectrum initial and final wavelength values together with the median wavelength step for each order and selected the maximum initial, minimum final, and minimum median step values as the values in the grid for each spectral order, a method also used by \citet{dumusque2018}.
We then interpolated each individual spectral order of each individual spectrum with cubic interpolation and co-added them using \verb+specutils+\footnote{\url{https://specutils.readthedocs.io/en/stable/index.html}} \citep{specutils2023}, which conveniently takes care of flux errors during these processes.

During the co-adding process we truncated the spectrum extremities of each spectral order by 10 \AA{} to avoid the drop in S/N at the edges due to the decrease in the blaze function flux, which will affect the normalisation process.
After co-adding all spectra, the resulting master spectrum was normalised to unity.
We used a two-step polynomial normalisation algorithm.
First we performed a sequential $\sigma$-clip to approximate the spectral order continuum by eliminating all flux values below the median minus 2$\sigma$ and fitted a fifth-order polynomial to the continuum.
Then, we repeated the process but this time using a sequential $\sigma$-clip with 1.7$\sigma$ and fitted the resulting continuum using a third-order polynomial.
The resulting continuum polynomial was then interpolated into the original wavelength grid and divided from the original flux.

\FloatBarrier
\section{Identification of spectral lines in the NIRPS spectra}\label{sec:line_id}

\subsection{Line identification using the Cretignier method}\label{sec:line_id_c20}

One of the important issues with line identification has to do with the definition of the line central bandpass.
Some line identification methods use a fixed bandpass in wavelength; however, this is not a good practice since the wavelength step is not constant in wavelength for most unprocessed spectra.
Fixing a bandpass value in RV space solves the wavelength step issue, but does not take into account the width of each line.
We were thus interested in a method that could take into account each line width.

\citet{cretignier2020} developed a method to identify and provide parameters such as line centre, depth and central bandpasses for symmetric spectral lines.
The method is based on the first and second derivatives of the spectrum whose maxima and minima provide constraints for the parameters and for the symmetry assessment.
Strong asymmetry in lines is assumed to be caused by line blending, and lines exhibiting asymmetry above a certain threshold are rejected by the selection criteria described in \citet[][Appendix C]{cretignier2020}.
The central bandpasses we  use to fit the line profiles and calculate the $pEW$ are the ones returned by this method, which we denominate as $\Delta\lambda_{C20}$.

To identify the spectral lines, We applied this method to the master spectrum of both stars.
Due to the overlapping of spectral orders our initial line list included duplicate lines at the orders extremities.
We identified these cases by selecting lines whose centres were closer than the central bandpass of one of the lines, and removed the line closer to the spectral order edges.
Since the S/N decreases towards the extremities due to the lower blaze function values in those regions, we selected lines from the duplicate list that were farther away from the wavelength initial and final values.
Our initial line search resulted in a list containing 6210 lines for Gl\,581 and 9396 for Proxima.
After applying the selection criteria discussed in \citet[][Appendix C]{cretignier2020}\footnote{We relaxed the continuum average parameter to 0.7 and did not use the number of maxima in the second derivative because we found that these constraints were removing wide and deep lines such as \ion{K}{i} $\lambda$12525.5, which could be well-fit and showed strong activity modulation in the line profile.} to this line list, we recovered a list of lines including 2391 for Gl\,581 and 3125 for Proxima.

\FloatBarrier

\subsection{Master spectra fitting and resulting line list}\label{sec:master_linelist}

To find the best profile model to fit to the observations and assess our fitting procedure, we fitted all lines we recovered with the Cretignier et al. method using the master spectrum as described in Sect. \ref{sec:line_fitting}.
The algorithm was incapable of properly fitting 12 of the Gl\,581 lines and three lines of the Proxima list.
Failure to fit identified lines can come from noise variability caused by low S/N for very shallow lines, residuals from telluric correction affecting the line profiles and/or blended lines that passed the selection criteria.
This resulted in a list of 2379 lines for Gl\,581 and 3122 lines for Proxima.

Some regions of the observations spectra have zero-valued flux counts that change from observation to observation.
These zero flux values are due to intense telluric lines with almost zero flux.
The pipeline forces them to zero to avoid spurious residuals.
Before fitting the observations, we excluded all lines with zero fluxes in the central bandpasses in at least one observation, excluding 58 lines from the Gl\,581 and 82 from the Proxima line lists.
Consequently, the final set of lines to be fitted to the observations includes 2321 lines for Gl\,581 and 3040 for Proxima.

\FloatBarrier

\section{Accuracy of the fitted observations parameters}\label{sec:acc_tests}

\begin{figure*}[t!]
    \centering
        \resizebox{0.9\hsize}{!}{\includegraphics{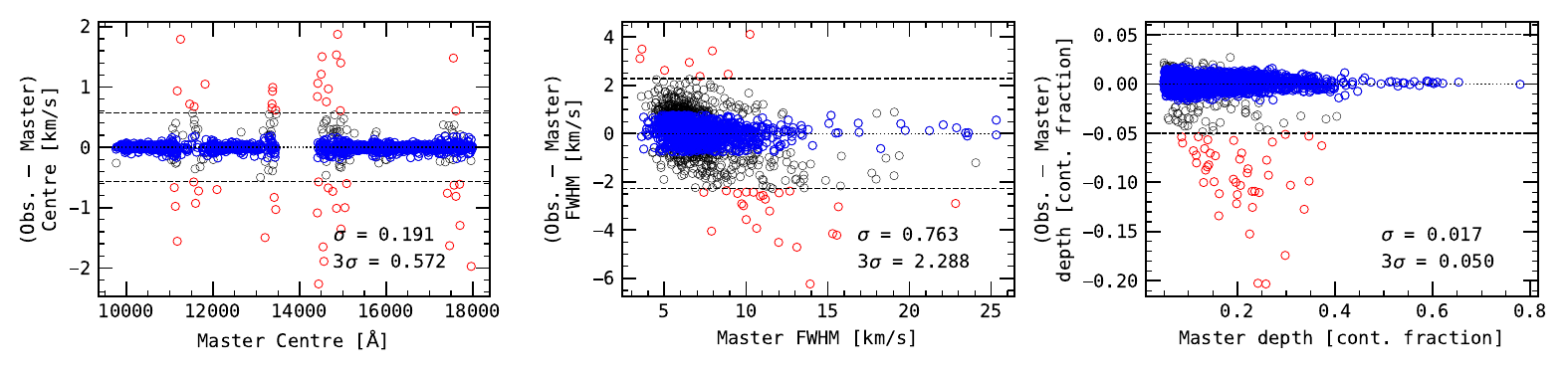}}
        \resizebox{0.9\hsize}{!}{\includegraphics{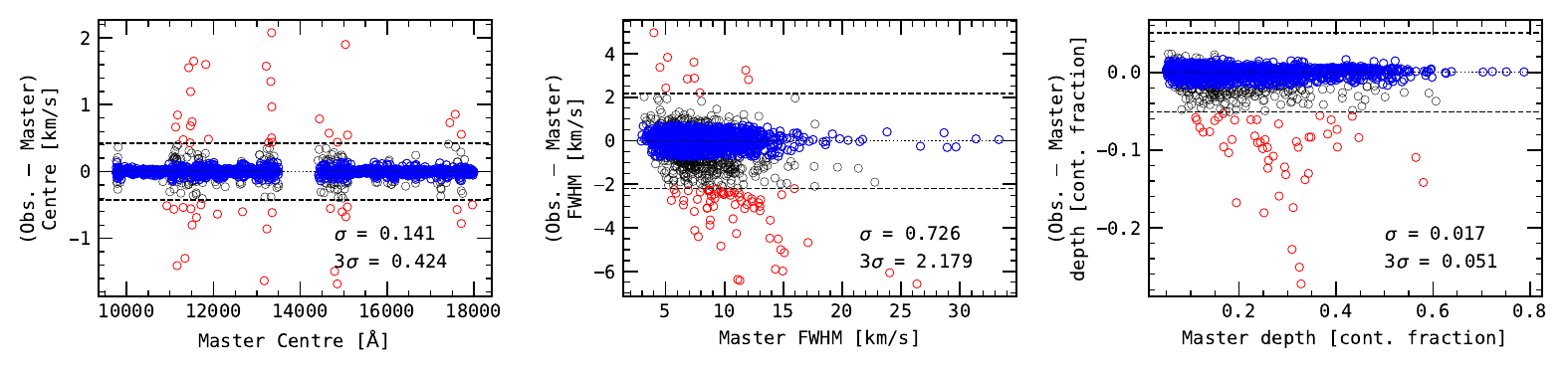}}

        \caption{Difference between the median parameter values extracted from the fit to observations and the master spectra parameters. The horizontal lines mark the position of the 3$\sigma$ levels around the median. The blue circles are lines with values below the 1$\sigma$ levels; the red circles are the removed lines with values above the 3$\sigma$ levels. Values of 1$\sigma$ and 3$\sigma$ levels are displayed for each parameter.}
        \label{fig:acc_tests}
\end{figure*}
\FloatBarrier

The accuracy of our observation fitting procedure was tested by comparing the median values of the time series from the observations fitting with those obtained from fitting the master spectrum to assess the accuracy of the time series.
The results are provided in Fig. \ref{fig:acc_tests}.
We are accurate to $\sim$200 m/s in line centre position, $\sim$750 m/s in FWHM, and $\sim$2\% in line depth, at the 1$\sigma$ level.
Lines with observation median parameters whose difference to the master values are above the 3$\sigma$ levels from the median were removed.
Most of these lines are located at the edges of the photometric bands $Y$, $J$ and $H$, as can be observed in the line centre panels.
This is due to the significant decrease in S/N at these regions as can also be observed in Fig. \ref{fig:NIRPS_master_orders}.
In terms of line depth, there is a tendency for the non-accurate lines to have lower depths as fitted from the observations than the master fit depths.

\section{Description of the CDS tables of the line lists of Gl\,581 and Proxima}

\begin{table*}[h!]
    \tiny
    \caption{Column descriptions for the tables of the line lists of Gl\,581 and Proxima. }
    \begin{center}
        \begin{tabular}{lcl}
            \hline\hline
            \multicolumn{1}{l}{Parameter} &
            \multicolumn{1}{c}{Units} &
            \multicolumn{1}{l}{Description} \\
            \hline
            \verb+ctr+ & $\mathrm{\AA}$ & Median of the fitted line centre time series. \\
            \verb+ctr_e+ & $\mathrm{\AA}$ & Median of the fitted line centre uncertainties time series. \\
            \verb+ctr_std+ & $\mathrm{\AA}$ & Standard deviation of the fitted line centre time series. \\
            \verb+fwhm+ & km/s & Median of the fitted FWHM time series. \\
            \verb+fwhm_e+ & km/s & Median of the fitted FWHM uncertainties time series. \\
            \verb+fwhm_std+ & km/s & Standard deviation of the fitted FWHM time series. \\
            \verb+depth+ & $1-F/F_c$ & Median of the fitted line depth time series. \\
            \verb+depth_e+ & $1-F/F_c$ & Median of the fitted line depth uncertainties time series. \\
            \verb+depth_std+ & $1-F/F_c$ & Standard deviation of the fitted line depth time series. \\

            \hline
            \verb+r_ctr_fwhm_nirps+ & $-$ & Pearson correlation coefficient between the line centre and the $FWHM_{CCF, NIRPS}$. \\
            \verb+pv_ctr_fwhm_nirps+ & $-$ & $p$--value of the correlation between the line centre and the $FWHM_{CCF, NIRPS}$. \\
            \verb+r_ctr_ha+ & $-$ & Pearson correlation coefficient between the line centre and the $I_{\mathrm{H}_{\alpha06}}$ index. \\
            \verb+pv_ctr_ha+ & $-$ & $p$--value of the correlation between the line centre and the $I_{\mathrm{H}_{\alpha06}}$ index. \\
            \verb+r_ctr_na+ & $-$ & Pearson correlation coefficient between the line centre and the $I_{NaI}$ index. \\
            \verb+pv_ctr_na+ & $-$ & $p$--value of the correlation between the line centre and the $I_{NaI}$ index. \\
            \verb+r_ctr_berv+ & $-$ & Correlation between line centre and BERV.\\

            \hline
            \verb+r_fwhm_fwhm_nirps+ & $-$ & Pearson correlation coefficient between the line FWHM and the $FWHM_{CCF, NIRPS}$. \\
            \verb+pv_fwhm_fwhm_nirps+ & $-$ & $p$--value of the correlation between the line FWHM and the $FWHM_{CCF, NIRPS}$. \\
            \verb+r_fwhm_ha+ & $-$ & Pearson correlation coefficient between the line FWHM and the $I_{\mathrm{H}_{\alpha06}}$ index. \\
            \verb+pv_fwhm_ha+ & $-$ & $p$--value of the correlation between the line FWHM and the $I_{\mathrm{H}_{\alpha06}}$ index. \\
            \verb+r_fwhm_na+ & $-$ & Pearson correlation coefficient between the line FWHM and the $I_{NaI}$ index. \\
            \verb+pv_fwhm_na+ & $-$ & $p$--value of the correlation between the line FWHM and the $I_{NaI}$ index. \\
            \verb+r_fwhm_berv+ & $-$ & Correlation between line FWHM and BERV.\\

            \hline
            \verb+r_depth_fwhm_nirps+ & $-$ & Pearson correlation coefficient between the line depth and the $FWHM_{CCF, NIRPS}$. \\
            \verb+pv_depth_fwhm_nirps+ & $-$ & $p$--value of the correlation between the line depth and the $FWHM_{CCF, NIRPS}$. \\
            \verb+r_depth_ha+ & $-$ & Pearson correlation coefficient between the line depth and the $I_{\mathrm{H}_{\alpha06}}$ index. \\
            \verb+pv_depth_ha+ & $-$ & $p$--value of the correlation between the line depth and the $I_{\mathrm{H}_{\alpha06}}$ index. \\
            \verb+r_depth_na+ & $-$ & Pearson correlation coefficient between the line depth and the $I_{NaI}$ index. \\
            \verb+pv_depth_na+ & $-$ & $p$--value of the correlation between the line depth and the $I_{NaI}$ index. \\
            \verb+r_depth_berv+ & $-$ & Correlation between line depth and BERV.\\

            \hline
            \verb+r_pew_act+ & $-$ & Pearson correlation coefficient between the best $pEW$ and the best activity index. \\
            \verb+pv_pew_act+ & $-$ & $p$--value of the correlation between the best $pEW$ and the best activity index. \\
            \verb+best_pew+ & $-$ & $pEW$ parameter responsible for the correlation in \verb+r_pew_act+. \\ 
            \verb+best_act+ & $-$ & Activity index responsible for the correlation in \verb+r_pew_act+. \\
            \verb+r_pew_berv+ & $-$ & Correlation between $pEW$ and BERV.\\
            \verb+bandpass+ & km/s & Central bandpass used to compute \verb+best_pew+. \\
            \verb+bandpass_c20+ & km/s & Central bandpass obtained from using \citet{cretignier2020} line identification method.\\

            \hline
            \verb+ctr_per+ & $d$ & Period detected using line centre time series.\\
            \verb+ctr_fap+ & $-$ & FAP of the period detected using line centre time series. \\
            \verb+fwhm_per+ & $d$ & Period detected using line FWHM time series.\\
            \verb+fwhm_fap+ & $-$ & FAP of the period detected using line FWHM time series. \\
            \verb+depth_per+ & $d$ & Period detected using line depth time series.\\
            \verb+depth_fap+ & $-$ & FAP of the period detected using line depth time series. \\
            \verb+pew_per+ & $d$ & Period detected using $pEW$ time series.\\
            \verb+pew_fap+ & $-$ & FAP of the period detected using line $pEW$ time series. \\
            \verb+best_pew+ & $-$ & $pEW$ parameter used in \verb+pew_per+. \\
            \hline
            flag & $-$ & 1 = correlation; 2 = rotation; 3 = correlation and rotation; 0 = no activity detection. \\
            \hline
        \end{tabular}
    \end{center}
    \tablefoot{All wavelength units are in a vacuum. This table is available at the CDS. The Flag parameter indicates if (1) a significant correlation was detected between at least one line parameter and one activity indicator; (2) a significant period consistent with the rotation period of the stars was detected; (3) both significant correlation and rotation period were detected; (0) No activity detection.}
    \label{tab:table_cds}
\end{table*}

\onecolumn
\section{Lines with significant rotation period detection for both stars}

\begin{table*}[h!]
    \centering
    \caption{Lines with significant rotation period detections common to  Gl\,581 and Proxima.\label{tab:prot_paper_table_common}}
    \small

    \begin{tabular}{cccccccccccc}
        \hline\hline
        \multicolumn{2}{c}{} &
        \multicolumn{5}{c}{Gl\,581} &
        \multicolumn{5}{c}{Proxima} \\
        \hline
        $\lambda_{0, med}$ & Sp. & $P$ & $FAP$ & $Param$ & $\langle depth \rangle$ & $\langle FWHM \rangle$ & $P$ & $FAP$ & $Param$ & $\langle depth \rangle$ & $\langle FWHM \rangle$ \\
        \multicolumn{1}{c}{[\text{\AA}]} &
        \multicolumn{1}{c}{} &
        \multicolumn{1}{c}{[d]} &
        \multicolumn{1}{c}{} &
        \multicolumn{1}{c}{} &
        \multicolumn{1}{c}{[$1-F/F_c$]} &
        \multicolumn{1}{c}{[km/s]} &
        \multicolumn{1}{c}{[d]} &
        \multicolumn{1}{c}{} &
        \multicolumn{1}{c}{} &
        \multicolumn{1}{c}{[$1-F/F_c$]} &
        \multicolumn{1}{c}{[km/s]} \\
        \hline
        9790.368 & \ion{Ti}{i} & 141.32 & 4e-09 & $FWHM$ & 0.66 & 8.6 & 79.24 & 8e-21 & $FWHM$ & 0.63 & 12.2 \\
        10433.372 & $-$ & 148.28 & 5e-05 & $FWHM$ & 0.17 & 6.0 & 78.41 & 2e-12 & $FWHM$ & 0.36 & 7.9 \\
        10908.741 & \ion{Cr}{i} & 138.20 & 2e-04 & $depth$ & 0.17 & 8.8 & 78.33 & 3e-06 & $pEW$ & 0.14 & 12.3 \\
        11641.442 & \ion{Fe}{i} & 140.26 & 6e-05 & $FWHM$ & 0.49 & 9.4 & 78.29 & 2e-18 & $FWHM$ & 0.49 & 14.2 \\
        11772.861 & \ion{K}{i} & 134.01 & 2e-09 & $pEW$ & 0.51 & 8.8 & 79.95 & 3e-34 & $FWHM$ & 0.72 & 21.5 \\
        11786.493 & \ion{Fe}{i} & 131.21 & 8e-04 & $pEW$ & 0.44 & 7.7 & 78.82 & 6e-24 & $depth$ & 0.37 & 10.1 \\
        11887.328 & \ion{Fe}{i} & 141.45 & 2e-06 & $pEW$ & 0.53 & 9.5 & 79.49 & 2e-14 & $FWHM$ & 0.54 & 13.7 \\
        12435.644 & \ion{K}{i} & 135.46 & 6e-12 & $FWHM$ & 0.54 & 9.5 & 81.42 & 2e-25 & $depth$ & 0.70 & 26.4 \\
        12525.528 & \ion{K}{i} & 140.79 & 4e-06 & $FWHM$ & 0.59 & 13.1 & 79.11 & 3e-26 & $FWHM$ & 0.75 & 31.5 \\
        12825.188 & \ion{Ti}{i} & 140.66 & 2e-07 & $FWHM$ & 0.48 & 8.7 & 79.21 & 3e-18 & $depth$ & 0.39 & 12.7 \\
        12903.288 & \ion{Mn}{i} & 138.33 & 6e-10 & $FWHM$ & 0.54 & 12.1 & 79.61 & 2e-26 & $depth$ & 0.52 & 17.2 \\
        13322.620 & \ion{Mn}{i} & 140.52 & 7e-06 & $FWHM$ & 0.46 & 10.5 & 79.61 & 4e-13 & $pEW$ & 0.43 & 15.4 \\
        15061.215 & \ion{Ca}{i} & 138.97 & 8e-04 & $FWHM$ & 0.24 & 7.9 & 78.99 & 3e-11 & $pEW$ & 0.21 & 16.4 \\
        15339.017 & \ion{Ti}{i} & 131.44 & 7e-04 & $centre$ & 0.34 & 12.6 & 78.78 & 3e-21 & $depth$ & 0.20 & 19.9 \\
        15792.876 & $-$ & 134.49 & 2e-04 & $centre$ & 0.05 & 4.0 & 92.20 & 1e-07 & $depth$ & 0.12 & 5.2 \\
        16062.960 & $-$ & 130.53 & 9e-04 & $depth$ & 0.17 & 6.1 & 96.21 & 5e-05 & $pEW$ & 0.27 & 6.7 \\
        17701.334 & $-$ & 146.26 & 8e-10 & $depth$ & 0.30 & 5.4 & 91.42 & 7e-05 & $pEW$ & 0.45 & 5.9 \\
        \hline
        \end{tabular}
    \tablefoot{$\lambda_{0, med}$ is the median of the line centres measured for both stars; Sp. is the species of the line obtained from the VALD3 database \citep{Ryabchikova2015}; $P$, $FAP$, $Param$, $\langle depth \rangle$, and $\langle FWHM \rangle$ correspond to the detected period, FAP for the detected period, line parameter responsible for the detected period, median line FWHM, and median line depth, respectively. All wavelength values are in vacuum.}
\end{table*}

\end{appendix}

\end{document}